\documentstyle[12pt]{article}
\begin{document}
\def\singlespacing{\baselineskip=12pt}
\def\doublespacing{\baselineskip=18pt}
\doublespacing

\noindent \today \\
Submitted to Phys.\ Rev.\ E

\bigskip
\bigskip

\begin{center}
\begin{large}
{\bf PHASE ORDERING KINETICS WITH EXTERNAL FIELDS AND BIASED INITIAL
     CONDITIONS} \\
\end{large}
\bigskip
\medskip
J. A. N. Filipe $^{(1,3)}$,  A. J. Bray $^{(1)}$ and Sanjay Puri $^{(2)}$ \\

\bigskip

(1) Theoretical Physics Group, Department of Physics and Astronomy,
The University, Manchester M13 9PL, UK \\
(2) School of Physical Sciences, Jawaharlal Nehru University,
New Delhi 110067, India. \\
(3) Present address: Physics Department, Brunel University, Uxbridge UB8
3PH, UK

\bigskip
\bigskip

{\bf ABSTRACT}
\end{center}
The late-time phase-ordering kinetics of the $O(n)$ model for a non-conserved
order parameter are considered for the case where the $O(n)$ symmetry is
broken by the initial conditions or by an external field. An approximate
theoretical approach, based on a `gaussian closure' scheme, is developed,
and results are obtained for the time-dependence of the mean order parameter,
the pair correlation function, the autocorrelation function, and the
density of topological defects [e.g. domain walls ($n=1$), or vortices
($n=2$)]. The results are in qualitative agreement with experiments on
nematic films and related numerical simulations on the two-dimensional
XY model with biased initial conditions.

\newpage

\pagestyle{plain}
\pagenumbering{arabic}

\section{INTRODUCTION}
\label{S:Intro}
The field of phase-ordering kinetics has seen a resurgence of interest in
recent years \cite{BrayRev}. It is by now well established that in the
late stages of ordering a scaling regime is entered, characterised by a
single time-dependent length scale $L(t)$. The existence of a single
characteristic scale has the consequence, for example, that the pair
correlation function for the order parameter field exhibits the scaling form
$C({\bf r},t) = f[r/L(t)]$, where $f(x)$ is a scaling function.
Systems with nonscalar order parameters have attracted special attention
recently, partly in response to experimental interest in liquid crystal
systems \cite{Yurke}. In particular, approximate methods
developed for calculating scaling functions for scalar fields
\cite{OJK:82,KYG:78}, have been extended to vector fields
\cite{BPT:91.92,LM:92:vectors,BH:92}.
So far, however, little effort has been devoted to the role of external
fields, or a symmetry-breaking bias in the initial conditions.

In this paper we study the phase ordering dynamics of systems with
non-conserved order parameter, in situations where an external field is
present and/or the initial state contains a bias.
We may call these systems `off-critical',
as opposed to critical systems whose initial state and dynamics are
rotationally invariant (or inversion symmetric for scalar fields). An
immediate difference relative to critical quenches is that the system will
(in most cases) reach global saturation exponentially fast, thus limiting
the duration of the scaling regime. We study both scalar and vector systems.
In the former there is a simple magnitude relaxation of the order parameter,
while in the latter there is also a slower orientational process.
Physical realizations are spin systems (magnets), nematic liquid crystals,
and order-disorder transitions in binary alloys.
The problem is addressed using an analytically solvable approach, the {\em
systematic approach} (SA) \cite{BH:93:SA}, which is a form of
`gaussian closure' approximation. At leading order, it reproduces the results
of earlier approximation schemes \cite{OJK:82,BPT:91.92}, but has the virtue
that it is (at least in principle) systematically improvable \cite{Zannetti}.
We derive an approximate linear equation for an auxiliary field $\vec{m}$.
This equation is exact in the limits $n\to\infty$
or $d\to\infty$, where $n$
and $d$ are the spin and space dimensions respectively.
{}From this we obtain predictions for the mean order parameter,
the topological defect density, and the pair correlation function.

The systematic approach was developed by Bray and Humayun (BH)
\cite{BH:93:SA} for simple non-conserved dynamics with no field or bias,
in which case the results of Ohta, Jasnow and Kawasaki (OJK) \cite{OJK:82},
and the extension thereof to vector fields by Bray, Puri and Toyoki (BPT)
\cite{BPT:91.92} are recovered in leading order. This approach has the
virtue, however, that it can in principle be systematically improved.
We are not concerned here with the possibility of improving the
leading order calculations, which is a problem of great technical
difficulty \cite{KB:93}, but in showing that the approach (in lowest order)
can be successfully extended to off-critical systems.

In section \ref{critical.SA} we discuss the original systematic approach.
For simplicity we begin with scalar fields. The extension to vector fields
is straightforward. In section \ref{off.SA} we generalize the scheme to
include a time-dependent external field and an initial bias.
In this case the relevant phenomenon is different for scalar and vector
systems and so the construction of the approach is also different.
For scalars the important effect of the field is a domain wall driving force,
while for vectors the bulk rotation of the order parameter plays an
important role.
The magnetization is compared with an earlier prediction from the large-$n$
solution of the time-dependent Ginzburg-Landau (TDGL) equation, by Bray and
Kissner (BK) \cite{BK:92}. We also compare the predictions for the planar
$XY$ model with an initial bias with data from experiment and simulations
on nematic liquid crystals \cite{PGY:94}.

\section{Critical Quenches}
\label{critical.SA}
An appropriate free-energy functional to describe a phase ordering system
is the Ginzburg-Landau form
\begin{equation}
F[\vec{\phi}] = \int d^dx\,\left[\frac{1}{2}\left(\nabla\vec{\phi}\right)^2
                  + V(\vec{\phi})\right]\ ,
\end{equation}
where $V(\vec{\phi})$ is a Mexican hat potential, e.g. $V(\vec{\phi}) =
(1-\vec{\phi}^2)^2$, with a degenerate ground state manifold
$\vec{\phi}^2 = 1$. For scalar systems, the potential has the usual
double-well structure, and the ground state has a discrete, two-fold
degeneracy.

An appropriate equation of motion for non-conserved dynamics is given by
the TDGL equation
\begin{equation}
\frac{\partial\vec{\phi}}{\partial t} = -\frac{\delta F}{\delta\vec{\phi}}
= \nabla^2\vec{\phi} - \frac{dV}{d\vec{\phi}}\ .
\label{TDGL}
\end{equation}
We will consider scalar and vector systems in turn.

\subsection{Scalar fields}
\label{critical.1}
According to the seminal work of Allen and Cahn \cite{AC:79}, the motion of
the interfaces in non-conserved systems is determined, in the absence of
a bulk driving force,  solely by their local
curvature. The detailed form of the potential $V(\phi)$ and the particular
distribution of initial conditions are not important to the late-stage
dynamics, and therefore to the scaling properties.
The only relevant feature of $V(\phi)$ is the double-well structure, which
generates and maintains well-defined interfaces, whilst its detailed shape
only affects the short-distance behaviour, such as the domain-wall profile.
Similarly, the details of the initial random configuration determine the
early-stage locations of the walls which, in the absence of any systematic
bias, should be irrelevant to the scaling properties. The systematic
approach exploits this freedom through a convenient choice of $V(\phi)$
and initial conditions.

The TDGL equation (\ref{TDGL}) for the evolution of a non-conserved, scalar
field $\phi({\bf r},t)$ reads (where primes indicate derivatives)
\begin{equation}
\frac{\partial\phi}{\partial t} = \nabla^2\phi-V'(\phi) \ .
\label{TDGL.scalar}
\end{equation}
In the `gaussian closure' schemes
\cite{BH:93:SA,OJK:82,BPT:91.92,M:89.90,BH:92,LM:92:vectors} a new
field $m({\bf r},t)$ is introduced, which varies smoothly on the domain
scale and whose zeros define the positions of the walls.
Following Mazenko \cite{M:89.90}, the transformation $\phi(m)$ is defined by
the equilibrium planar profile of an isolated wall, which satisfies
\begin{equation}
\phi''(m) = V'(\phi) \ , \label{mazenko.map.1.II}
\end{equation}
with boundary conditions
\begin{equation}
\phi(\pm\infty) = \pm 1 \ \ , \ \ \phi(0) = 0 \ . \label{boundaries.1}
\end{equation}
With this choice, near a wall (where this can be regarded as flat) $m$ plays
the role of a distance from the wall, whilst inside the domains $m$ is
typically of order $L(t)$. Rewriting equation (\ref{TDGL.scalar}) in terms of
$m$, and using (\ref{mazenko.map.1.II}) to eliminate $V'$, gives
\begin{equation}
\dot{m} = \nabla^2 m-\frac{\phi''(m)}{\phi'(m)}\left(1-(\nabla m)^2\right) \ .
\label{m.eq.1}
\end{equation}
The non-linear profile function $\phi(m)$ depends on the potential $V(\phi)$,
so for general potentials equation (\ref{m.eq.1}) offers no obvious
simplification over the original TDGL equation (\ref{TDGL.scalar}).
As argued above, however, the scaling functions should be insensitive to the
potential, and thus to the details of the profile function.
The results of the OJK \cite{OJK:82} and Mazenko \cite{M:89.90} approaches,
for instance, make no explicit reference to the potential.
The key step in the present approach is to exploit this idea, by choosing a
particular form of $V(\phi)$ such that equation (\ref{m.eq.1}) is simplified.

Bray and Humayun (BH) \cite{BH:93:SA} chose the profile function to satisfy
\begin{equation}
\phi''(m) = - m\,\phi'(m) \ , \label{SA.condition}
\end{equation}
which is equivalent, via (\ref{mazenko.map.1.II}), to a particular form of
the potential, as shown below.
Integrating (\ref{SA.condition}) with the boundary conditions
(\ref{boundaries.1}) gives the wall profile function
\begin{eqnarray}
\phi(m) & = & \sqrt{2/\pi}\,\int_0^m dx\,\exp\left(-x^2/2\right) \nonumber \\
      & = & {\rm erf}\left(m/\sqrt{2}\right) \ , \label{SA.profile}
\end{eqnarray}
where ${\rm erf}(x)$ is the error function.
On the other hand, integrating (\ref{mazenko.map.1.II}) once, with the
zero of the potential defined by $V(\pm 1)=0$, gives
\begin{eqnarray}
V(\phi) & = & (1/2)\,\left(\phi'\right)^2 \ = \ (1/\pi)\,\exp(-m^2)
\label{V.phi} \\
 & = & (1/\pi)\,\exp\left(-2\left\{{\rm erf}^{-1}(\phi)\right\}^2\right)\ ,
\label{SA.potential}
\end{eqnarray}
where ${\rm erf}^{-1}(x)$ is the inverse error function.
Note that (\ref{SA.profile}) is a monotonic mapping, with the constraint
$\phi^2\leq 1$ due to the boundary conditions (\ref{boundaries.1}). It follows
that (\ref{SA.potential}) only determines $V(\phi)$ within the region
$\phi^2\leq 1$.
For instances, we have $V(\phi)\simeq 1/\pi-\phi^2/2$ for $\phi^2\ll 1$, while
$V(\phi)\simeq (1/4)(1-\phi^2)^2\,|\ln(1-\phi^2)|$ for $\phi^2\simeq 1$.
At a temperature of $T=0$, however, if the initial condition satisfies
$\phi^2({\bf r},0)\leq 1$ everywhere, the equation of motion
(\ref{TDGL.scalar}) implies that $\phi^2({\bf r},t)\leq 1$ everywhere
at any later time. So $\phi({\bf r},t)$ does not depend
on $V(\phi)$ for $\phi^2> 1$, and there is no need to know the potential in
this region (if $T>0$, however, the points $\phi=\pm 1$ must be global minima
of $V(\phi)$, in order to ensure stability against thermal fluctuations).

With the choice (\ref{SA.condition}), equation (\ref{m.eq.1}) reduces to
\begin{equation}
\dot{m} = \nabla^2 m + \left(1-(\nabla m)^2\right)\,m \ . \label{m.eq.2}
\end{equation}
This equation, though still non-linear, is much simpler than the original TDGL
equation, while retaining the physical ingredients essential to describe
the universal scaling properties.
To make further progress, and to recover the usual OJK results, one simply
replaces $\left(\nabla m\right)^2$ by its average (over the initial conditions)
\begin{equation}
\left(\nabla m\right)^2 \,\rightarrow\, \left<\left(\nabla m\right)^2\right>
\ .
\label{linearize}
\end{equation}
Then (\ref{m.eq.2}) becomes the self-consistent linear equation for
$m({\bf r},t)$:
\begin{equation}
\dot{m} = \nabla^2 m + a(t)\,m \ , \label{m.eq.3}
\end{equation}
with
\begin{equation}
a(t) = 1 - \left<(\nabla m)^2\right> \ . \label{at.1}
\end{equation}

As in OJK's theory, to proceed with the calculations one takes the initial
conditions for $m$ to be gaussian distributed, with zero mean and correlator
(in Fourier space)
\begin{equation}
\left<m_{{\bf k}}(0)m_{{\bf k}'}(0)\right> =
\Delta (2\pi)^d \delta({\bf k}+{\bf k}') \ , \label{SA.IC}
\end{equation}
representing short-range spatial correlations at $t=0$.
Hence, from (\ref{m.eq.3}), $m({\bf r},t)$ is a gaussian field at all times.
The formal solution of (\ref{m.eq.3}) in Fourier space is
\begin{equation}
m_{{\bf k}}(t) = m_{{\bf k}}(0)\, \exp\left(-k^2t+b(t)\right) \ , \label{mk.1}
\end{equation}
with the definition
\begin{equation}
b(t) = \int_0^t dt'\, a(t') \ . \label{bt.1}
\end{equation}
The function $b(t)$ is determined self-consistently, as follows.
Inserting (\ref{mk.1})-(\ref{bt.1}) into definition (\ref{at.1}) gives
\begin{equation}
a(t) = db/dt = 1 -
\Delta\,\exp(2b(t))\,\int\frac{d^dk}{(2\pi)^d}k^2 \exp\left(-2k^2t\right) \ .
\label{at.2}
\end{equation}
Although this equation can be readily solved for the whole time range after
the quench, it suffices for our purposes to extract the large-$t$ behaviour of
$b(t)$ which is most easily obtained directly from (\ref{at.2}).
In the scaling limit we expect that $a(t)\to 0$ (in fact we expect that
$a(t)\sim 1/t$ if all terms in (\ref{m.eq.3}) scale the same way).
Neglecting the left-hand side of (\ref{at.2}) and performing the momentum
integral, gives
\begin{equation}
\exp(b(t)) \simeq \left(t/t_0\right)^{\frac{d+2}{4}} \ \ \ , \ \ \ t\gg t_0 \ ,
\label{bt.2}
\end{equation}
with the definition
\begin{equation}
t_0^{\frac{d+2}{2}} = \frac{d\Delta}{4(8\pi)^{d/2}} \ . \label{t0}
\end{equation}
{}From (\ref{bt.2}) we immediately get
\begin{equation}
a(t) \ \simeq \ \frac{d+2}{4t} \ \ \ , \ \ \ t\gg t_0 \ , \label{at.3}
\end{equation}
confirming the consistency of the previous assumption.
The short-time cut-off $t_0$ was introduced as a device to prevent the
breakdown of the momentum integral in (\ref{at.2}) as $t\to 0$. It is
equivalent to imposing a large-$k$ cut-off (of order $1/t_0^{1/2}$)
associated with the smallest length in the system (e.g.\ the defect core
size $\zeta$), which removes the mathematical singularities. So $t_0$
should be small. The condition $t\gg t_0$ characterizes the times for
which the above asymptotic forms are valid, i.e.\ it defines the scaling
regime ($L(t)\gg\zeta$) in which we are interested.
Throughout this paper we will, however, for computational convenience,
take $t$ down to $t_0$ while still using (\ref{bt.2}).
This implies that the calculated scaling functions will neither be accurate
nor universal in the early-time regime $t\sim t_0$, but will not affect
their behaviour in the true scaling regime.
In this regime, the scaling functions will not depend on the specific
way in which the cut-off has been introduced.

Having determined $b(t)$, we may use (\ref{bt.2}) to write the explicit
result for $m_{{\bf k}}(t)$, valid for large-$t$,
\begin{equation}
m_{{\bf k}}(t) = m_{{\bf k}}(0)\,\left(t/t_0\right)^{\frac{d+2}{4}}
\exp(-k^2t) \ , \label{mk.2}
\end{equation}
from which the two-time correlator, in Fourier and real space, follows
immediately:
\begin{equation}
\left<m_{{\bf k}}(t_1)m_{-{\bf k}}(t_2)\right> =
\Delta\,\left(t_1t_2/t_0^2\right)^{\frac{d+2}{2}}
\exp\left(-k^2(t_1+t_2)\right) \ , \nonumber
\end{equation}
and
\begin{equation}
C_0(1,2) \equiv \left<m(1)m(2)\right> = \frac{4\sqrt{t_1t_2}}{d}\,
\left(\frac{4t_1t_2}{(t_1+t_2)^2}\right)^{d/4}
\exp\left(-\frac{r^2}{4(t_1+t_2)}\right) \ , \label{SA.C0}
\end{equation}
where `1' and `2' are the usual shorthand for space-time points
$({\bf r}_1,t_1)$ and $({\bf r}_2,t_2)$, and $r=|{\bf r}_1-{\bf r}_2|$.
{}From this, we find the expected scaling form
\begin{equation}
S_0(t) \equiv \left<m^2\right> = \frac{4t}{d} \sim L(t)^2 \ , \label{SA.S0}
\end{equation}
in agreement with the physical interpretation of $m$ as a length.
It is interesting that the linear term $a(t)m$ in (\ref{m.eq.3}),
with $a(t)$ given by (\ref{at.3}), exactly reproduces the Oono-Puri
extension of the OJK theory \cite{OP:88}, designed to eliminate the implicit
time-dependence of the interface-width, which is unphysical. The resulting
difference in the time factors of (\ref{SA.C0}) and (\ref{SA.S0}), however,
does not affect the scaling
functions, which as usual only depend on the normalized correlator
\begin{equation}
\gamma(1,2) \equiv
\frac{\left<m(1)m(2)\right>}{\sqrt{\left<m(1)^2\right>\left<m(2)^2\right>}}
= \left(\frac{4t_1t_2}{(t_1+t_2)^2}\right)^{d/4}
\exp\left(-\frac{r^2}{4(t_1+t_2)}\right) \ . \label{SA.gamma.t1t2}
\end{equation}

Finally, we evaluate the correlation function of the field $\phi$.
Since from (\ref{SA.S0}) $m$ is typically of order $\sqrt{t}$ at late-times, it
follows from (\ref{SA.profile}) that the field $\phi$ saturates ($\phi=\pm 1$)
almost everywhere at late-times.
As a consequence, we can make the usual simplification $\phi(m)={\rm sgn}(m)$
as far as scaling properties are concerned.
The calculation of the average
$C(1,2) = \left<{\rm sgn}(m(1)){\rm sgn}(m(2))\right>$ for a gaussian field
$m({\bf r},t)$ is the same as in the OJK theory (see the Appendix).
Therefore, the OJK scaling function $C(12) = (2/\pi) \sin^{-1}[\gamma(1,2)]$
is recovered.

We conclude this subsection with a technical remark on the above procedure.
To make the replacement (\ref{linearize}) in a controlled way, BH systematize
the treatment by attaching to the field $m$ an internal `color' index $\alpha$,
running from 1 to $N$, and generalizing (\ref{m.eq.2}) to
\begin{equation}
\dot{m_{\alpha}} = \nabla^2 m_{\alpha} +
\left(1-(1/N)\sum_{\beta=1}^N \left(\nabla m_{\beta}\right)^2\right)\,
m_{\alpha} \ . \label{m.eq.4}
\end{equation}
Equation (\ref{m.eq.2}) is the case $N=1$. In the limit $N\to\infty$ the
`color` sum is replaced by its average, giving the linear equation
(\ref{m.eq.3}) (with $m$ standing for one of the $m_{\alpha}$).
This procedure allows, in principle, for a systematic treatment of the
results in powers of $1/N$.
The replacement (\ref{linearize}) is also justified in the limit $d\to\infty$,
when $\left(\nabla m\right)^2$ is also a sum of a large number of random
variables with relative fluctuations $\sim 1/\sqrt{d}$.
This scheme, though less simple to systematize \cite{BH:93:SA},
makes clear that the leading order results become exact for large-$d$.
We stress, however, that a we are not concerned here with the calculation
of $O(1/N)$ corrections to the previous results. Our purpose is to
extend the leading order calculations to off-critical systems.

\subsection{Vector fields}
\label{critical.n}
For non-conserved vector fields, the starting point is the TDGL equation in
the form (\ref{TDGL}). This time a {\em vector} field $\vec{m}({\bf r},t)$ is
introduced, whose zeros define the locations of the topological defect cores.
The transformation $\vec{\phi}(\vec{m})$ is now defined by an equilibrium
solution of (\ref{TDGL}), which satisfies the vector analogue of
(\ref{mazenko.map.1.II}) \cite{BH:92,LM:92:vectors}
\begin{equation}
\nabla^2_m\vec{\phi} = dV/d\vec{\phi} \ , \label{mazenko.map.n}
\end{equation}
where $\nabla^2_m=\sum^n_{\alpha=1}\partial^2/\partial m_{\alpha}^2$ is the
Laplacian in $\vec{m}$ space.
We look for a radially symmetric solution of (\ref{mazenko.map.n}):
\begin{equation}
\vec{\phi}(\vec{m}) = g(\rho)\,\hat{m} \ , \label{SA.radial}
\end{equation}
where $\rho\equiv|\vec{m}|$ and $\hat{m}\equiv\vec{m}/|\vec{m}|$,
with boundary conditions
\begin{equation}
g(\infty) = 1 \ \ \ , \ \ \ g(0) = 0 \ , \label{boundaries.n}
\end{equation}
(though any solution differing from (\ref{SA.radial}) by a global rotation
is equally acceptable).
Thus, $g(\rho)$ is the equilibrium profile function for a topological defect
in the field $\vec{\phi}$, with $|\vec{m}|$ representing the distance from
the defect core \cite{BH:92,LM:92:vectors}.
A solution of this form makes sense, of course, only for $n\leq d$ when
singular topological defects exist.
Rewriting the TDGL equation (\ref{TDGL}) for each component of $\vec{\phi}$
in terms of $\vec{m}$, and using (\ref{mazenko.map.n}) to eliminate
$dV/d\phi_a$, gives
\begin{equation}
\sum_{b}  \frac{\partial\phi_a}{\partial m_b} \frac{\partial m_b}{\partial t}
= \sum_{b}  \frac{\partial\phi_a}{\partial m_b} \nabla^2 m_b +
\sum_{bc} \frac{\partial^2\phi_a}{\partial m_b\partial m_c}
\nabla m_b\!\cdot\!\nabla m_c -  \nabla^2_m \phi_a \ . \label{vm.eq.1}
\end{equation}
As in the scalar case, in order to establish a linear equation for $\vec{m}$
one replaces the quadratic factor $\nabla m_i\cdot\nabla m_j$ by its average
(over the initial conditions):
\begin{equation}
\nabla m_b\!\cdot\!\nabla m_c \,\rightarrow\,
\left<\left(\nabla m\right)^2\right>\delta_{bc} \ . \label{linearize.n}
\end{equation}
Here we used the fact that $\left<(\nabla m_b)^2\right>$ is independent of $b$,
from global isotropy, to write it as $\left<(\nabla m)^2\right>$ where $m$ is
any component of $\vec{m}$.
As before, one can attach an additional `colour' index $\alpha$ ($=1,...,N$)
to the vector $\vec{m}$, such that the replacement (\ref{linearize.n})
corresponds to taking the limit $N\to\infty$ in the theory. In this case,
however, it also corresponds to the limit $n\to\infty$.
With (\ref{linearize.n}), (\ref{vm.eq.1}) simplifies to
\begin{equation}
\sum_{b}  \frac{\partial\phi_a}{\partial m_b} \frac{\partial m_b}{\partial t}
= \sum_{b}  \frac{\partial\phi_a}{\partial m_b} \nabla^2 m_b -
\left(1-\left<(\nabla m)^2\right>\right)\,\nabla^2_m \phi_a \ . \label{vm.eq.2}
\end{equation}
Finally, one would like to eliminate the explicit dependence of this equation
on $\vec{\phi}(\vec{m})$.
Once again, BH exploit the expected insensitivity of the scaling functions to
the details of the potential, by choosing the function $\vec{\phi}(\vec{m})$
to satisfy
\begin{equation}
\nabla^2_m \vec{\phi} \ = \
- \left(\vec{m}\!\cdot\!\vec{\nabla}_m\right)\,\vec{\phi} \ = \
- \sum_b m_b\,\frac{\partial\vec{\phi}}{\partial m_b} \ ,
\label{SA.condition.n}
\end{equation}
a direct generalization of (\ref{SA.condition}).
Substituting the radially symmetric form (\ref{SA.radial}) in
(\ref{SA.condition.n}) gives
\begin{equation}
g'' + \left(\frac{n-1}{\rho}+\rho\right) g'-\frac{n-1}{\rho^2} g \ = \ 0 \ ,
\label{SA.profile.n}
\end{equation}
the equation for the profile function $g(\rho)$ with boundary conditions
(\ref{boundaries.n}), which generalizes (\ref{SA.condition}).
For small-$\rho$ the solution is linear
\begin{equation}
g(\rho) \sim \rho \ \ \ \ \ \ (\rho\to 0) \ , \label{g.small-x}
\end{equation}
while for large-$\rho$ the profile saturates as
\begin{equation}
g(\rho) \ \simeq \ 1-(n-1)/2\rho^2 \ \ \ \ \ \ (\rho\to\infty) \ ,
\label{g.large-x}
\end{equation}
from which we can take $\zeta^2=(n-1)/2$ as a definition of the defect core
size.
The potential $V(\phi)$ corresponding to this choice of the profile function
can be deduced from (\ref{mazenko.map.n}), though it seems unlikely that a
closed-form expression can be derived.

With the choice (\ref{SA.condition.n}), equation (\ref{vm.eq.2}) now becomes
\begin{equation}
\left( \frac{\partial\vec{m}}{\partial t}-\nabla^2\vec{m}-a(t)\vec{m} \right)
\!\cdot\!\vec{\nabla}_m\phi_a \ = \ 0 \ , \label{vm.eq.3}
\end{equation}
where $a(t)$ is given by (\ref{at.1}). Let us define the vector
\begin{equation}
\vec{\Omega} \ \equiv \
\partial\vec{m}/\partial t - \nabla^2 \vec{m} - a(t) \vec{m} \ . \label{Omega}
\end{equation}
In principle, equation (\ref{vm.eq.3}) allows for solutions where
$\vec{\Omega}$ and $\vec{\nabla}_m\phi_a$ are orthogonal vectors without
$\vec{\Omega}$ being zero. However, inserting the radial form
(\ref{SA.radial}) in (\ref{vm.eq.3}) gives
\begin{equation}
\left(\vec{\Omega}\!\cdot\!\vec{\nabla}_m\right)\:\vec{\phi} \ = \
\frac{g(\rho)}{\rho}\,\vec{\Omega}_\perp + g'(\rho)\,\vec{\Omega}_\parallel \
= \ 0 \ , \label{vm.eq.4}
\end{equation}
where $\vec{\Omega}_\perp$ and $\vec{\Omega}_\parallel$ are the components of
$\vec{\Omega}$ perpendicular and parallel to $\hat{m}$, i.e.
\begin{eqnarray}
\vec{\Omega}_\perp & = &
\vec{\Omega}-\left(\vec{\Omega}\!\cdot\!\hat{m}\right)\hat{m} \nonumber \\
\vec{\Omega}_\parallel & = &
\left(\vec{\Omega}\!\cdot\!\hat{m}\right)\hat{m} \ . \label{Omega.comps}
\end{eqnarray}
The coefficients $g'$ and $g/\rho$ in (\ref{vm.eq.4}) only vanish for
$\rho\to\infty$, so the orthogonal vectors $\vec{\Omega}_\perp$ and
$\vec{\Omega}_\parallel$ must vanish separately everywhere.
Therefore $\vec{\Omega}=0$ is the only physical solution of (\ref{vm.eq.3}).
Then equation (\ref{vm.eq.1}) reduces to the self-consistent linear form
(\ref{m.eq.3})-(\ref{at.1}), holding separately for each component of
$\vec{m}$. Taking gaussian initial conditions for each component, with zero
mean and correlator (\ref{SA.IC}), the usual procedure leads to the same
asymptotic form (\ref{SA.C0}) for the correlator $\left<m_a(1)m_a(2)\right>$
($a=1,...,n$), and thus to the same form (\ref{SA.gamma.t1t2}) for the
normalized correlator
\begin{equation}
\gamma(1,2) \equiv \frac{\left<\vec{m}(1)\!\cdot\!\vec{m}(2)\right>}
{\sqrt{\left<\vec{m}(1)^2\right>\left<\vec{m}(2)^2\right>}} \ . \label{gamma.n}
\end{equation}

The final step is to evaluate the pair correlation function,
$C(1,2)=\left<\vec{\phi}(1)\!\cdot\!\vec{\phi}(2)\right>$.
This proceeds exactly as in the BPT treatment \cite{BPT:91.92}:
from (\ref{SA.S0}) $\rho$ scales as $\sqrt{t}$, then according to
(\ref{SA.radial})--(\ref{boundaries.n}) we can replace the profile $g(\rho)$
by 1 and the function $\vec{\phi}(\vec{m})$ by $\hat{m}$ at late-times.
Since from (\ref{m.eq.3}) the components of $\vec{m}$ are gaussian fields
at all times, the calculation of the average
$C(1,2) = \left<\hat{m}(1)\!\cdot\!\hat{m}(2)\right>$ leads to the standard
BPT scaling function \cite{BPT:91.92} (see also the appendix):
\begin{equation}
C(1,2) = \frac{n\gamma}{2\pi}\left[B\left(\frac{n+1}{2},\frac{1}{2}\right)
\right]^2\,_2F_1\left[\frac{1}{2},\frac{1}{2};\frac{n+2}{2};
\gamma(1,2)^2\right]\ ,\label{BPT}
\end{equation}
where $B(x,y)$ is the beta function and $_2F_1[a,b;c;z]$ is the
Hypergeometric function.

Remarks similar to those of the previous section, regarding the possibility
of systematically improving the results by expanding in $1/N$, apply here.
In addition, the present (leading order) results are expected to become
exact both for large-$d$ and for large-$n$, when the linearization of the
$\vec{m}$ equation, leading to the gaussian property of $\vec{m}$, is
justified.

\section{Off-Critical Quenches}
\label{off.SA}
In the last section we discussed in detail the systematic approach
for the $O(n)$ model representing a system undergoing a `critical quench'.
A remarkable feature of this approach is that it can be extended in a
simple manner to treat the situation when an external field is present
and/or the initial state contains a bias.

Regarding the external field, the key point of the approach is to select a
convenient effective driving force which simplifies the equation for
$\vec{m}({\bf r},t)$ and which captures the essential physics of the problem.
To achieve this, we focus on the relevant effect of the external field in the
asymptotic dynamics of the ordering system, which is different for scalar and
vector systems. The initial bias, on the other hand, only enters the equation
for $\vec{m}$ through the initial condition.
To leading order in the systematic approach, we obtain a linear equation for
the auxiliary field similar to the critical case, but with an extra driving
force $h$ for $n=1$ and $\sqrt{\left<m^2\right>}\,\vec{h}$ for $n>1$.
To evaluate the expectation values associated with the order parameter
(including $\left<\vec{\phi}\right>$, which was zero in the critical case),
we extend the gaussian calculations for a distribution centered about a
non-zero average. In particular, we derive the off-critical extension of the
BPT scaling function (\ref{BPT}).

\subsection{Scalar fields}
\label{off.1}

Our starting point for a non-conserved, scalar system in the presence of
an external field, is the usual TDGL equation (\ref{TDGL.scalar}), but now
the potential $V(\phi)$ is asymmetric.

The energy density due to an external field (e.g.\ a magnetic field) can be
modelled by a linear term coupling the field to the order parameter, which
biases the order parameter in the direction of the field. Hence the potential
$V(\phi)$ has the form
\begin{equation}
V(\phi;h)=V_0(\phi) - h(t)\,\phi \ , \label{V.linear}
\end{equation}
where $V_0(\phi)$ is the usual symmetric, double-well potential, and $h(t)$
is a general time-dependent external field. For a spin system, for example,
the above linear form only holds if $\phi$ is far
from saturation and $h$ is not too large, otherwise the system will respond
non-linearly to the field and terms such as $O(h\phi^3)$ become important.
In the present continuous order parameter system, the saturation values of
$\phi$ simply get modified by $h$.
For positive $h$ and given $t$, the potential $V(\phi;h)$ has an asymmetric
double-well form, with the right- (left-) hand minimum lower (higher) than
in $V_0(\phi)$.
For our purposes, it is convenient to rewrite the potential in the
generic form
\begin{equation}
V(\phi;h) = V_0(\phi) - h(t)\,V_1(\phi) \ , \label{V.asym}
\end{equation}
where $V_1(\phi)$ is a monotonic, odd function of $\phi$.
Then the evolution equation for the order parameter reads
\begin{equation}
\frac{\partial\phi}{\partial t} = \nabla^2\phi - V_0'(\phi) + h(t)\,V_1'(\phi)
\ .\label{TDGL.asym}
\end{equation}

As in the treatment of the case $h=0$, we can exploit the insensitivity of the
domain growth to specific details of the potential by choosing an especially
convenient form for $V(\phi)$. This relies on the physical idea that the
interface motion only depends on the local curvature $K$ and on the local
field $h$. To motivate this idea, we derive the generalized version of
Allen-Cahn's equation for the interface motion, valid for an arbitrary
potential. If $g$ is a coordinate normal to the interface (increasing
in the direction of increasing $\phi$), the TDGL equation (\ref{TDGL.scalar})
can be recast near an interface as
\begin{equation}
-\left(\frac{\partial\phi}{\partial g}\right)_t
\left(\frac{\partial g}{\partial t}\right)_\phi =
K\left(\frac{\partial\phi}{\partial g}\right)_t
+ \left(\frac{\partial^2\phi}{\partial g^2}\right)_t - V'(\phi)\ ,
\end{equation}
where the curvature $K$ is the divergence of the `outward' normal (i.e.
pointing in the direction of increasing $\phi$) to the interface.
Multiplying through by $(\partial\phi/\partial g)_t$ and integrating over $g$
through the interface gives the local velocity of the interface,
$v=(\partial g/\partial t)_\phi$ as
\begin{equation}
v = - K + \Delta V/\Sigma \ ,
\label{AC.asym}
\end{equation}
where $\Delta V$ is the change in the potential $V(\phi,h)$ across the
interface, and $\Sigma = \int dg(\partial \phi/\partial g)_t^2$ is the
surface tension. The essential point here, is that the interface motion
depends on the external field only through $\Delta V$.
So, as far as domain growth is concerned, all that matters is the difference
between the minima of the potential $V(\phi;h)$ for the two bulk phases.
This gives us a great deal of flexibility in the choice of $V_1(\phi)$.
In particular, we may choose the minima of $V(\phi;h)$ to remain at
$\phi=\pm 1$, avoiding inessential contributions to the magnetization from
`stretching' the field.

As usual, we introduce the auxiliary variable $m({\bf r},t)$ whose zeros
determine the wall positions. Rewriting equation (\ref{TDGL.asym}) in terms of
$m$ yields
\begin{equation}
\phi'\dot{m} = \phi'\nabla^2 m +
\phi''(\nabla m)^2 - V_0'(\phi) + h(t) V_1'(\phi) \ . \label{hm.eq.1}
\end{equation}
We define the transformation $\phi=\phi(m)$ by the equilibrium wall-profile
for the asymmetric potential. This situation corresponds to a planar moving
wall (moving, say, in the positive $x$-direction) with position $x_0(t)$ and
velocity $v(t)=\dot{x}_0(t)$.
So, the equilibrium solution is of the form
\begin{equation}
\phi({\bf r},t) = \phi(x-x_0(t)) \ . \label{moving.wall}
\end{equation}
Substituting this in (\ref{TDGL.asym}) gives the equation for the steady-state
profile
\begin{eqnarray}
V'(\phi;h) & = & \phi''(m) + v(t)\,\phi'(m) \label{asym.map} \\
         & = & V_0'(\phi) - h(t) V_1'(\phi) \nonumber \ ,
\end{eqnarray}
which generalizes (\ref{mazenko.map.1.II}), and has the same boundary
conditions (\ref{boundaries.1}). This immediately suggests the convenient
choices
\begin{eqnarray}
V_0'(\phi) &=& \phi''(m)    \label{V0.map} \\
V_1'(\phi) &=& \phi'(m)  \ , \label{V1.map}
\end{eqnarray}
along with the identification
\begin{equation}
v(t) = - h(t) \ . \label{v.h}
\end{equation}
Note that (\ref{V0.map}) determines the profile for a given $V_0(\phi)$, or
vice-versa, while (\ref{V1.map}) fixes $V_1(\phi)$ for a given profile
$\phi(m)$. In addition, we adopt the BH choice (\ref{SA.condition}) for the
function $\phi(m)$, i.e.\ $\phi'' = - m\,\phi'$.
Inserting (\ref{V0.map}), (\ref{V1.map}) and (\ref{SA.condition})
in (\ref{hm.eq.1}), we obtain the simpler equation
\begin{equation}
\dot{m} = \nabla^2 m + \left(1-(\nabla m)^2\right)\, m + h(t) \ .
\label{hm.eq.2}
\end{equation}

The condition (\ref{SA.condition}) gives once again the error function profile
(\ref{SA.profile}), and amounts, via (\ref{V0.map}), to choosing the previous
form (\ref{SA.potential}) for the symmetric potential $V_0(\phi)$.
The potential $V_1(\phi)$ is obtained by integrating (\ref{V1.map}),
with boundary condition $V_1(0)=0$. This gives
\begin{eqnarray}
V_1(\phi) & = & \int_0^m dx\,\left(\phi'(x)\right)^2 \ = \
\frac{2}{\pi} \int_0^m dx\,\exp\left(-x^2\right) \nonumber \\
& = & \frac{1}{\sqrt{\pi}} {\rm erf}(m)                \label{V1.m} \\
& = & \frac{1}{\sqrt{\pi}} {\rm erf}\left(\sqrt{2}{\rm erf}^{-1}(\phi)\right)
\ . \label{V1.phi}
\end{eqnarray}
where we used (\ref{SA.profile}). Alternatively, from (\ref{V1.map}) and
(\ref{V.phi}), we get
\begin{eqnarray}
V_1(\phi) & = & \int_0^{\phi} dx\,\sqrt{2V_0(x)} \label{V1.V0}
\end{eqnarray}
Again, (\ref{V1.phi}) or (\ref{V1.V0}) only defines $V_1(\phi)$ for
$\phi^2 \le 1$, but this is the only region we require for $T=0$.
In particular, we have $V_1(\phi)\simeq \sqrt{2/\pi}\,\phi$ for $\phi^2\ll 1$,
while $V_1(\phi)\simeq \pm \left[1/\sqrt{\pi} -
(1/4)(1-\phi^2)^2\sqrt{|\ln(1-\phi^2)|/2}\right]$ for $\phi^2\simeq 1$.
Therefore, $V_1(\phi)$ has a sigmoid shape, with linear behaviour
(\ref{V.linear}) for small-$\phi$, while saturating to $\pm 1/\sqrt{\pi}$
in the bulk and so acting only on the interfaces.
We can relate our field $h(t)$ to an effective field driving the interface
dynamics by matching the differences $\Delta V = - 2h_{eff}$, which follows
from (\ref{V.linear}), and $\Delta V = - h \Delta V_1 = - 2h/\sqrt{\pi}$.
This yields: $h_{eff}=h/\sqrt{\pi}$.
Note that if we had kept the form (\ref{V.linear}) throughout ($V_1'=1$), the
minima of $V(\phi;h)$ would be shifted by the field $h$, which is inconsistent
with the use of the profile function (\ref{SA.profile}), where $\phi^2\leq 1$.
This would also give a non-linear term $h/\phi'\sim\exp(m^2/2)$ in
(\ref{hm.eq.2}), amplifying the effect of the field in the bulk (and thus
shifting the minima of $V$) and destroying the mathematical simplicity of the
equations.

\bigskip

\noindent
{\bf Expectation values} \\
\noindent
In order to solve equation (\ref{hm.eq.2}), we use the replacement
(\ref{linearize}) for $\left(\nabla m\right)^2$, which corresponds to taking
the limit $d\to\infty$ or a large number of `colours', $N$.
With this, (\ref{hm.eq.2}) becomes the self-consistent equation for
$m({\bf r},t)$:
\begin{equation}
\dot{m} = \nabla^2 m + a(t)\,m + h(t) \ , \label{hm.eq.3}
\end{equation}
with $a(t)$ given by (\ref{at.1}); a simple extension of (\ref{m.eq.3}).
In addition, we allow for a uniform bias in the initial state, taking gaussian
initial conditions for $m$ with non-zero mean, and only short-ranged
correlations
\begin{eqnarray}
\left<m({\bf r},0)\right> & = & m_0   \label{m0} \\
\left<m({\bf r},0)m({\bf 0},0)\right>_c & \equiv &
\left<m({\bf r},0)m({\bf 0},0)\right> - \left<m({\bf r},0)\right>^2 =
\Delta \delta({\bf r}) \ . \label{mm0}
\end{eqnarray}
The Fourier components of $m({\bf r},0)$ still satisfy
$\left<m_{{\bf k}}(0)\right>=0$ and (\ref{SA.IC}) for ${\bf k}\neq 0$.

Solving equation (\ref{hm.eq.3}) in Fourier space, gives
\begin{equation}
m_{{\bf k}}(t) = m_{{\bf k}}(0)\, \exp\left(-k^2t+b(t)\right) +
(2\pi)^d\delta(k)\,\int_0^t dt'\, h(t')\,
\frac{\exp(b(t))}{\exp(b(t'))} \ , \label{hmk.eq.1}
\end{equation}
with $b(t)$ defined by (\ref{bt.1}).
Since $\left<(\nabla m)^2\right>$ is unaffected by the presence of a
uniform bias and field, it is the same as for $h=0=m_0$. Hence, the
function $b(t)$ is still determined by the self-consistency equation
(\ref{at.2}), leading once again to the asymptotic form (\ref{bt.2})
for times $t\gg t_0$, with $t_0$ given by (\ref{t0}). Inserting this
result in (\ref{hmk.eq.1}) yields, for large-$t$,
\begin{equation}
m_{{\bf k}}(t) = \left(t/t_0\right)^{\frac{d+2}{4}} \left\{
m_{{\bf k}}(0)\,\exp\left(-k^2t\right)  + (2\pi)^d\delta({\bf k})\, \eta(t)
\right\} \ , \label{hmk.eq.2}
\end{equation}
with
\begin{equation}
\eta(t) = \int_{t_0}^t dt'\,h(t')\,
          \left(\frac{t_0}{t'}\right)^{\frac{d+2}{4}} \ , \label{eta}
\end{equation}
where the lower cut-off $t_0$ was introduced into the integral to
account for the breakdown of the form used for $b(t)$ at short-times.
Although the form used for $b(t)$ is only valid in the scaling regime
($t\gg t_0$) we shall be taking $t\geq t_0$ for computational
convenience. This gives an inaccurate description of the initial transient
behaviour prior to the scaling regime, but is irrelevant for the scaling
behaviour (for more details see the discussion following (\ref{at.3})).

Averaging (\ref{hmk.eq.2}) over initial conditions and taking the Fourier
transform, gives the average value of $m({\bf r},t)$ at late-times
\begin{equation}
\left<m(t)\right> = \left(t/t_0\right)^{\frac{d+2}{4}}
\left[m_0 + \eta(t)\right] \ .\label{M0}
\end{equation}
{}From (\ref{hmk.eq.2}) we also obtain the previous expressions, (\ref{SA.C0})
and (\ref{SA.S0}), for the second cumulants of $m$, i.e.\ the connected pair
correlator and one-point correlator
\begin{eqnarray}
C_0(1,2) \equiv \left<m(1)m(2)\right>_c & = &
\left<m(1)m(2)\right>-\left<m(1)\right>\left<m(2)\right> \ , \label{C0c} \\
S_0(t) \equiv \left<m^2\right>_c & = & \left<m^2\right> - \left<m\right>^2 \ .
\label{S0c}
\end{eqnarray}
This immediately gives the same form (\ref{SA.gamma.t1t2}) for the
off-critical normalized correlator
\begin{equation}
\gamma(1,2) \ \equiv \
\frac{\left<m(1)m(2)\right>_c}{\sqrt{\left<m(1)^2\right>_c
\left<m(2)^2\right>_c}}\ = \ \left(\frac{4t_1t_2}{(t_1+t_2)^2}\right)^{d/4}
\exp\left(-\frac{r^2}{4(t_1+t_2)}\right) \ . \label{SA.gammac.t1t2}
\end{equation}
For equal-times ($t_1=t_2=t$), (\ref{SA.gammac.t1t2}) simplifies to
\begin{equation}
\gamma(1,2) = \exp\left(-\frac{r^2}{8t}\right) \ . \label{SA.gammac.t}
\end{equation}

As usual, at late-times we can replace the profile function $\phi(m)$
(eq.\ (\ref{SA.profile})) by ${\rm sgn}(m)$ to
evaluate the expectation values associated with the field $\phi$.
According to (\ref{hm.eq.3}), $m({\bf r},t)$ is a gaussian field at all times,
with probability distribution defined by the non-zero mean (\ref{M0})
and the second cumulants (\ref{C0c})-(\ref{S0c}).
Evaluating the gaussian expectation values we obtain scaling functions with
arguments (\ref{M0}), (\ref{S0c}) and (\ref{SA.gammac.t1t2}).
The details of these calculations are given in the Appendix,
so we shall state the results only.

\bigskip

\noindent
{\bf Magnetization} \\
\noindent
For the average value of the order parameter $\phi$, i.e.\ the magnetization,
we obtain
\begin{equation}
\left<\phi(t)\right> \ = \ \left<{\rm sgn}(m)\right> \ = \
{\rm erf}\left(M(t)\right) \ , \label{phi}
\end{equation}
where $M(t)$ is a dimensionless average of $m$:
\begin{eqnarray}
M(t) & \equiv & \frac{\left<m\right>}{2\sqrt{\left<m^2\right>_c}} \label{M} \\
& = & \sqrt{d/8t_0}\,
\left(\frac{t}{t_0}\right)^{d/4}\,\left[m_0 + \eta(t)\right] \ . \label{SA.M}
\end{eqnarray}
Some results for $\left<\phi(t)\right>$ are presented in Figures 1 and 2.
The result using the profile (\ref{SA.profile}) is $\left<\phi\right>=
{\rm erf}\left[M/(1+\zeta^2/2\left<m^2\right>_c)^{1/2}\right]$,
with $\zeta=\sqrt{2}$ the wall-width for this profile.
Clearly this reduces to (\ref{phi}) in the scaling regime.
In the saturation limit ($|M|\to\infty$) we expand the error function in
(\ref{phi}) to give
\begin{equation}
\left<\phi(t)\right> \ \simeq \ {\rm sgn}(M(t))
\left[1-\frac{\exp\left(-M^2(t)\right)}{|M(t)|\sqrt{\pi}}\right]
\ \ \ , \ \ \ |M(t)|\gg 1 \ , \label{phi.large-M}
\end{equation}
and so, the magnetization saturates exponentially fast.
Conversely, in the `initial-growth' regime when $M$ is small (e.g.\ for
small $m_0$ and $h(t)$ and $t$ not too large), (\ref{phi}) reduces to
\begin{equation}
\left<\phi(t)\right> \ \simeq \ (2/\sqrt{\pi})\, M(t) \ \ \ , \ \ \
|M(t)|\ll 1 \ , \label{phi.small-M}
\end{equation}
showing that in this limit $M$ plays the role of a magnetization.

The time-dependence of $\left<\phi\right>$ is determined by the spatial
dimension $d$. On one hand there is the prefactor $t^{d/4}$ in $M$. On the
other hand note that $M$ has two contributions with different physical
origins: one comes from the initial bias while the other is due to the
external field. Which term dominates at late-times ($t\gg t_0$) depends on
the value of $d$ and on the explicit form of $h(t)$.
It is also interesting to note that if we set $h(t)$ to vanish over a time
$t_1>t_0$, then
\begin{equation}
M(t) = \left(t/t_1\right)^{d/4}\, M(t_1) \ , \label{recurrence}
\end{equation}
i.e.\ the `magnetization' induced at subsequent times grows in the same manner
as the one induced by the initial bias.
To illustrate the nature of the interplay between the magnetization growth
induced by the initial bias and that driven by the external field, let us
consider a time-independent field $h$. In this case (\ref{eta}) reduces to
\begin{eqnarray}
\eta(t) & = & h\,\int_{t_0}^t dt'\,
          \left(t_0/t'\right)^{\frac{d+2}{4}} \label{eta.h} \\
& = & A\,ht_0 \left\{\left(t/t_0\right)^{\frac{2-d}{4}}-1\right\}
    \ \ \sim \ \ h\left(t/t_0\right)^{\frac{2-d}{4}} \ , \ \ d<2
\label{373} \\
& = & ht_0\:\ln(t/t_0) \ \ \ \ \ \ \ \ \ \ \ \ \ \ \ \ \ \ \ \ , \ \ d=2
\nonumber \\
& = & A\,ht_0 \left\{1-\left(t_0/t\right)^{\frac{d-2}{4}}\right\}
    \ \ \sim \ \ h \ \ \ \ , \ \ d>2 \ . \nonumber
\end{eqnarray}
where $A=4/|d-2|$.
The asymptotic forms on the right stand for late-times $t\gg t_0$.
For $d\leq 2$ the integral (\ref{eta.h}) is dominated by times of order $t$,
so the continued influence of the field dominates over the initial bias in
(\ref{SA.M}) at late-times.
For $d>2$ the integral is dominated by short-times,
so $\eta(t)$ scales like the initial bias $m_0$ in (\ref{SA.M}). This means
that the prevalent effect of the field is the growth induced at early-times,
a similar situation to that described by (\ref{recurrence}).
For a general time-dependent field $h(t)$ the crossover between these two
regimes will be different, but clearly the factor $1/t'^{(d+2)/4}$ in
(\ref{eta}) diminishes the effect of the field as time progresses.
In particular, a sinusoidal field is less effective in reversing the
magnetization during later cycles. This situation is illustrated in Figure 1.

Using linear response and scaling arguments BK \cite{BK:92} predicted the
scaling form of the magnetization in the `initial growth' regime when
$\left<\phi\right>\ll 1$. Our expression (\ref{phi.small-M})-(\ref{eta.h}) for
$\left<\phi\right>$ agrees closely with their scaling form. The essential
modification is that $t^{d/4}=L^{d/2}$ gets replaced by $L^{\lambda}$, where
$\bar{\lambda} = d-\lambda $ is the exponent characterizing the decay
of the autocorrelation function.
Hence, the crossover between the two regimes no longer occurs at
$d=2$ but at the dimension where $\lambda=1$ \cite{BK:92}.
In the systematic approach, as in the OJK theory, the exponent $\lambda$ is
given by
\begin{equation}
\lambda_{SA} = d/2 \ . \label{SA.lambda}
\end{equation}
The BK scaling prediction provides a means of measuring $\lambda$
without having to measure correlations between fields at different times.
The result (\ref{phi}), although approximate, provides an explicit expression
for the magnetization, describing the complete time-range from the early
scaling regime to final saturation ($\left<\phi\right>\to\pm 1$).

To conclude this discussion, we address the situation when the initial bias
$m_0$ and the constant field $h$ have opposite signs. According to
(\ref{373}) and (\ref{SA.M}) the magnetization is reversed after a time
$t_{rev}$ given by
\begin{eqnarray}
\left(t_{rev}/t_0\right)^{\frac{2-d}{4}} = 1 + |m_0|/|ht_0\,A|
\ \ \ \ , \ \ {\rm for} \ \ d<2  \ , \label{tinv.1} \\
t_{rev}/t_0 = \exp\left(|m_0|/|ht_0|\right)
\ \ \ \ , \ \ {\rm for} \ \ d=2 \ . \nonumber
\end{eqnarray}
For $d>2$, however, a sufficiently weak field, $|h|<|m_0|/At_0$, can never
overpower the initial bias and the system evolves towards the metastable state
corresponding to $\left<\phi\right>={\rm sgn}(m_0)$.
This result is for a system at zero temperature (there is no noise in
the equation of motion). At temperatures $T>0$ thermal fluctuations will
eventually drive the system towards stable equilibrium (corresponding to
$\left<\phi\right>={\rm sgn}(h)$) for any $d$.
Does the above statement agree with what we should expect for $T=0$?
We recall the two effects of the magnetic field. Initially it creates a bias
which grows in the same manner as the bulk magnetization induced by $m_0$
(eq.\ (\ref{recurrence})); the net result is the sum of these two.
In addition it produces a wall driving force which adds to the wall curvature
force (which exists for $d>1$) \cite{footnote}.
The net force dictates which of the two effects dominates at late-times.
Within our approach the curvature dominates for $d>2$ (but not for $d\leq 2$),
in which case the magnetization is determined by the net bias growth, i.e.\
the balance between $m_0$ and $h$. We would expect the curvature to dominate
for any $d>1$ when the field $h$ is small enough, but we cannot say precisely,
within this qualitative argument, how small $h$ should be compared to $m_0$.
The results for $1<d\leq 2$ are probably an artifact of the approach, which is
in fact a large-$d$ treatment. Therefore, the behaviour found for $d>2$ is
probably the only physical behaviour when $d>1$ and $h$ is very small.

\bigskip

\noindent
{\bf Pair correlations} \\
\noindent
For the pair correlation function of the field $\phi$, we obtain (see the
Appendix)
\begin{eqnarray}
C(1,2) & = & \left<{\rm sgn}(m(1)){\rm sgn}(m(2))\right> \nonumber \\
& = & \frac{2}{\pi}\: \int_0^{\gamma(1,2)}
\frac{dy}{\sqrt{1-y^2}}\:\exp\left(-\Gamma(y)\right) \ + \
\left<\phi(1)\right>\!\left<\phi(2)\right> \ , \label{SAh.C}
\end{eqnarray}
with
\begin{equation}
\Gamma(y) = \frac{M^2(1)+M^2(2)-2yM(1)M(2)}{1-y^2} \ , \label{Gamma.y}
\end{equation}
where $M(i)=M(t_i)$ is given by (\ref{SA.M}) and $\gamma(1,2)$ is
given by (\ref{SA.gammac.t1t2}).
In the critical limit ($m_0=h=0$) we have $M=\left<\phi\right>=0$,
hence the usual OJK form $C = (2/\pi) \sin^{-1}\gamma$ is recovered.
For widely-separated times ($t_2/t_1\to\infty$) or for infinitely separated
points ($r\to\infty$), we have $\gamma\to 0$, and so
$C(1,2)\to\left<\phi(1)\right>\left<\phi(2)\right>$.

At equal-times ($M(1)=M(2)=M(t)$), (\ref{SAh.C}) reduces to
\begin{eqnarray}
\label{E:1}
C(x,M) & = & \frac{2}{\pi}\: \int_0^{\gamma(x)} \frac{dy}{\sqrt{1-y^2}}\:
\exp\left(\frac{-2M^2}{1+y}\right) \ + \ \left<\phi\right>^2  \\
 & = & 1 - \frac{2}{\pi}\: \int_{\gamma(x)}^1 \frac{dy}{\sqrt{1-y^2}}\:
\exp\left(\frac{-2M^2}{1+y}\right)\ ,
\label{SAh.C.x}
\end{eqnarray}
where $x=r/t^{1/2}$ is the scaling variable and $\gamma(x)$ given by
(\ref{SA.gammac.t}).
It is worth studying the limiting behaviour of this expression.
For $\gamma\to 1$ ($x\to 0$) one obtains, to leading order in $x$,
\begin{equation}
C(x,M) = 1 - \exp\left(-M^2(t)\right)\:\left[(1/\pi)x+O(x^3)\right]
\ \ , \ \ x\to 0 \ . \label{SAh.C.small-x}
\end{equation}
Note, that the term $\exp(-M^2(t))\,x$ is, up to constants, $\rho_{def}(t)\,r$,
where $\rho_{def}$ is the wall density, given by (\ref{SAh.rho.m}) below.
This shows that the correlation function $C(x,M)$ has a small-distance
Porod's regime, as expected.
In the limit $\gamma\to 0$ ($x\to\infty$), we take $r\to\infty$ keeping $t$
fixed in order to have $M(t)$ fixed. Then, Taylor expanding (\ref{E:1})
we obtain
\begin{equation}
C(x,M) = \frac{2}{\pi}\:\exp\left(-2M^2(t)\right)\:\left(\gamma +
M(t)^2\gamma^2 + ...\right) \ + \ \left<\phi\right>^2 \ \ , \ \ x\to\infty \ .
\label{SAh.C.large-x}
\end{equation}
In the saturation limit, when $M(t)^2\to\infty$ and $\left<\phi\right>^2\to 1$,
$C(x,t)$ saturates exponentially to 1 in (\ref{SAh.C.x})-(\ref{SAh.C.large-x}).
The equal-time pair correlation scaling function $C(x,M)$
is shown in Figure 3, for a constant external field.

\bigskip

\noindent
{\bf Density of walls} \\
\noindent
We are also interested in the average density of wall `area' per
unit volume of the system, $\rho_{def}(t)$.
There are a number of ways of calculating $\rho_{def}$ within
the gaussian approximation. For example,
$\rho_{def} = \left<\left(\nabla\phi\right)^2\right>/\Sigma =
- \left(\nabla^2 C\right)_{r=\zeta}/\Sigma$, where $\Sigma$ is the surface
tension,  can be evaluated from
(\ref{SAh.C.x}). The results differ by a constant amplitude, as a
consequence of the inaccuracy of the approximation, but our main
interest is to obtain the time-dependence of the wall density.
A simple procedure \cite{B:93:defects,LM:92:defects} which easily generalizes
for vector fields, is to write the wall density at each point ${\bf r}$
in terms of the auxiliary field $m$, as
\begin{equation}
\rho_{def}({\bf r},t) = \delta(m({\bf r},t))\,|\nabla m({\bf r},t)| \ ,
\label{rho.m}
\end{equation}
where $|\nabla m|$ is the Jacobian between $m$ and the spatial coordinate
${\bf r}$. If we adopt the usual definition of $m$ as a normal
coordinate near defects, then $|\nabla m|=1$ at defects, giving simply
$\rho({\bf r},t)=\delta(m({\bf r},t))$ \cite{B:93:defects}
(note that this is true for the exact $m$, but not for the gaussian $m$,
for which $|\nabla m|$ fluctuates). Now we make the
gaussian approximation for $m$, with the one-point distribution function
\begin{equation}
P(m) = \frac{1}{ \sqrt{2\pi S_0} }\,
\exp\left( \frac{-(m-\left<m\right>)^2}{2S_0} \right) \ . \label{Pm}
\end{equation}
Then, the average wall density is given by
\begin{eqnarray}
\rho_{def}(t) & = & \left<\delta(m)\right> \label{aver.rho.m} \ = \ P(0)
\nonumber \\
& = & \sqrt{\frac{d}{8\pi}}\:\frac{\exp\left(-M^2(t)\right)}{\sqrt{t}} \ ,
\label{SAh.rho.m}
\end{eqnarray}
where we used (\ref{SA.S0}), and $M(t)$ is given by (\ref{SA.M}). This
result agrees, up to constants, with an earlier prediction by OJK
\cite{OJK:82}. When $m_0=h=0$, (\ref{SAh.rho.m}) gives the usual scaling
form $\rho\sim L^{d-1}/L^d \sim 1/L$.
The ordering process is much faster in the presence of an external field
or initial bias, hence the exponential decay of the wall density.

\subsection{Vector fields}
\label{off.n}

The evolution of a non-conserved vector system in the presence of an
external field is given by the TDGL equation (\ref{TDGL}). The usual form
of the potential $V(\vec{\phi})$ (for 'soft spins') is a direct
generalization of (\ref{V.linear}), i.e.\
\begin{equation}
V(\vec{\phi};\vec{h})=V_0(\vec{\phi}) - \vec{h}(t)\!\cdot\!\vec{\phi} \ ,
\label{V.linear.vector}
\end{equation}
where $V_0(\vec{\phi})$ is the usual `Mexican hat' potential, with a
symmetric ground-state manifold $\vec{\phi}^2=1$, and $\vec{h}(t)$ is a
general time-dependent external field.
The full potential $V(\vec{\phi};\vec{h})$ has an asymmetric (tilted)
`Mexican hat' shape, with a single global minimum where $\vec{\phi}$ is
parallel to $\vec{h}$ and has maximum length (among the all minima), and a
metastable state where $\vec{\phi}$ is antiparallel to $\vec{h}$ and has
minimum length.
These saturation lengths depend on the value of $|\vec{h}|$.
For the remaining directions of $\vec{\phi}$ there is only a minimum along
each direction, corresponding to a local saturation length, which varies
continuously between the two above values.

As usual for vector systems, we want to introduce a vector auxiliary field
$\vec{m}$ whose zeros define the locations of the topological defect cores.
By analogy with the previous cases, in order to define the transformation
$\vec{\phi}(\vec{m})$ let us imagine how a steady-state configuration of the
field $\vec{\phi}$ looks like in the presence of a slowly varying field
$\vec{h}$.
For definiteness consider a system with $n=d$, which has point topological
defects (i.e.\ vortices for $d=2$ and monopoles for $d=3$), along with
anti-defects. These attract each other and eventually annihilate at late-times.
Without an external field the defects are radially symmetric: at the defect
core $\vec{\phi}$ has zero length and undefined direction, while away from
the defect core $\vec{\phi}$ rapidly saturates in length. The effect of the
field $\vec{h}$ is to deform the defects: all `force lines' (except one)
emerging from the defect core tend to line up with $\vec{h}$ away from the
core. Close enough to the defect core, however, the field $\vec{\phi}$ will
still be isotropic,
so we may still use the `radially symmetric' form (\ref{SA.radial}), i.e.\
\begin{equation}
\vec{\phi}(\vec{m}) = g(\rho)\,\hat{m} \ , \label{SA.radial.II}
\end{equation}
where $\rho=|\vec{m}|$ and $g(\rho)$ is a function yet to be determined.
In the vicinity of a core, where $\hat{m}$ is isotropic, $\rho$ and $g(\rho)$
play the same role as in the case $\vec{h}=0$, while in the bulk the direction
of $\hat{m}$ is (on average) parallel to $\vec{h}$. Note that, by symmetry,
there is no direct force on the defect core due to the external field.

Contrary to the scalar case, we conclude that an important effect of the
external field occurs in the bulk, where the relevant phenomenon is the
orientational ordering of $\vec{\phi}$ along the direction of $\vec{h}$.
In the bulk we expect $\vec{\phi}$ to rapidly relax to its saturation length,
i.e.\ to the `local' minimum of $V(\vec{\phi};\vec{h})$ which depends on the
local angle between $\vec{\phi}$ and $\vec{h}$.
In other words, since field `stretching' is a local process while field
rotation involves long-range correlations, the longitudinal effect of
$\vec{h}$ (along $\vec{\phi}$) should be instantaneous as compared to
transverse effects.
The same assumption has motivated large-$n$ studies of this problem using a
`hard-spin' model \cite{PFT:87,BK:92}. This leads us to interpret $g(\rho)$
in (\ref{SA.radial.II}) as the {\em stationary} profile function of a
defect in the field $\vec{\phi}$, with $|\vec{m}|$ representing the distance
from the defect along a (time-dependent) `force line' defined by $\hat{m}$.
Hence, as far as bulk rotation and the slow defect dynamics are concerned,
we may once again (but for reasons different from the scalar case) ignore
the inessential `stretching' of the field, by choosing the local minima of
$V(\vec{\phi};\vec{h})$ to remain at $\vec{\phi}^2=1$. This amounts to taking
the same boundary conditions (\ref{boundaries.n}) for $g(\rho)$.

With these insights, let us rewrite the TDGL equation (\ref{TDGL}) as
\begin{equation}
\frac{\partial\vec{\phi}}{\partial t} =
\nabla^2\vec{\phi}-\frac{dV_0(\vec{\phi})}{d\vec{\phi}} +
\vec{U}(\vec{\phi};\vec{h}) \ , \label{TDGLh.vector}
\end{equation}
where $\vec{U}(\vec{\phi};\vec{h})$ is the driving force due to the field
$\vec{h}$ (which would be simply $\vec{h}$ with the potential
(\ref{V.linear.vector})).
As usual, we expect the scaling functions to be insensitive to the details of
the potential, or equivalently, of the driving force. These should only
affect phenomena on the scale of the defect core size $\zeta$, but not on the
scale of the bulk characteristic length $L(t)$. Thus, we can make convenient
choices for $V_0$ and $\vec{U}$ in order to simplify the equation of motion
for $\vec{m}$.

As in the case $\vec{h}=0$, we define the relation between the function
$\vec{\phi}(\vec{m})$ and the potential $V_0(\vec{\phi})$ by the equation
(\ref{mazenko.map.n}).
Rewriting (\ref{TDGLh.vector}) in terms of $\vec{m}$ gives (\ref{vm.eq.1}) but
with an extra term $U_a$. As usual, we use the replacement (\ref{linearize.n})
which is justified in the large-$n$ limit, where
$\left<(\nabla m)^2\right>$ still stands for any component of $\vec{m}$.
Note that $(\nabla m)^2$ is unaffected by a {\em uniform} external field or
bias, so the system is isotropic as far as the average
$\left<(\nabla m)^2\right>$ is concerned.
We also choose the function $\vec{\phi}(\vec{m})$ to satisfy
(\ref{SA.condition.n}).
This leads once again, via (\ref{SA.radial.II}), to the equation
(\ref{SA.profile.n}) for the profile function $g(\rho)$, with small and
large-$\rho$ behaviour (\ref{g.small-x})-(\ref{g.large-x}).
It also corresponds to the same choice for the symmetric potential
$V_0(\vec{\phi})$ as in the case $\vec{h}=0$.
Putting it all together and using definition (\ref{Omega}), the TDGL
equation (\ref{TDGLh.vector}) then becomes
\begin{equation}
\left(\vec{\Omega}\!\cdot\!\vec{\nabla}_m\right)\,\vec{\phi} =
\vec{U}(\vec{\phi};\vec{h}) \ . \label{vmh.eq.1}
\end{equation}
The key step now, is to rewrite the driving force in the anisotropic form:
\begin{equation}
\vec{U}(\vec{\phi};\vec{h}) = U_\perp(g)\,\vec{h}_\perp +
U_\parallel(g)\,\vec{h}_\parallel \ , \label{U.comps}
\end{equation}
and to use (\ref{SA.radial.II}) to write the left-hand side of
(\ref{vmh.eq.1}) as in (\ref{vm.eq.4}), i.e.\ as
\begin{equation}
\left(\vec{\Omega}\!\cdot\!\vec{\nabla}_m\right)\:\vec{\phi} =
\frac{g(\rho)}{\rho}\,\vec{\Omega}_\perp + g'(\rho)\,\vec{\Omega}_\parallel \ ,
\label{vmh.eq.LHS}
\end{equation}
where the indices $_\parallel$ and $\perp$ refer to the components of the
vectors parallel and perpendicular to $\vec{\phi}$
(eq.\ (\ref{Omega.comps})).
Matching (\ref{U.comps}) and (\ref{vmh.eq.LHS}) gives immediately
\begin{eqnarray}
\Omega_\perp(\vec{m}) & = & \frac{\rho}{g(\rho)}\,h_\perp\,U_\perp \ ,
\nonumber \\
\Omega_\parallel(\vec{m}) & = & \frac{1}{g'(\rho)}\,h_\parallel\,U_\parallel
\ . \label{vmh.eq.2}
\end{eqnarray}
To obtain a simple, `isotropic' equation of motion for $\vec{m}$, we adopt
the following choice for the driving force:
\begin{eqnarray}
U_\perp & = & g(\rho) \ , \nonumber \\
U_\parallel & = & \rho\,g'(\rho) \ , \label{U.choice}
\end{eqnarray}
This corresponds, via (\ref{U.comps}), (\ref{g.small-x}) and (\ref{g.large-x}),
to having
\begin{eqnarray}
\vec{U} &\sim& \rho\,\vec{h} \ \ \ \ \ \ (\rho\to 0) \ , \nonumber \\
\vec{U} &\simeq& \vec{h}_\perp \ + \
(1-\vec{\phi}^2)\,\left(\vec{h}_\parallel-\vec{h}_\perp/2\right)
\ \ \ \ \ \ (\rho\to\infty) \ , \label{U.limits}
\end{eqnarray}
in total agreement with the our requirements from the above discussion:
$\vec{U}$ vanishes at the defect cores, while being isotropic in their
vicinity; in the bulk $\vec{U}$ reduces to its transverse component,
forcing $\vec{\phi}$ to rotate and to line up with $\vec{h}$.
Just as in the scalar case, the driving force derives from a potential:
$\vec{U}(\vec{\phi};\vec{h})=dV_1(\vec{\phi};\vec{h})/d\vec{\phi}$.
To get an equation for $V_1$ one writes $\vec{U}$ in terms of
$\vec{\phi}_\parallel$ and $\vec{\phi}_\perp$, the components of
$\vec{\phi}$ parallel and perpendicular to $\vec{h}$.
However, according to (\ref{U.comps}) and (\ref{U.choice}) one cannot deduce
$V_1$ without solving equation (\ref{SA.profile.n}) for $g(\rho)$.
Nonetheless, all that matters is the driving force which is physically
correct.

\bigskip

\noindent
{\bf Expectation values} \\
\noindent
We now solve the equation for $\vec{m}$ and obtain its expectation values.
With the choice (\ref{U.choice}) and using definition (\ref{Omega}) of
$\vec{\Omega}$, the equation of motion (\ref{vmh.eq.2}) becomes
\begin{equation}
\partial \vec{m}/\partial t = \nabla^2\vec{m} + a(t)\,\vec{m} +
\rho\,\vec{h}(t) \ . \label{vmh.eq.3}
\end{equation}
In the spirit of the systematic approach, in order to obtain a linear
equation for $\vec{m}$ we replace $\vec{m}^2$ by its average in $\rho$:
\begin{equation}
\rho = \sqrt{\vec{m}^2} \rightarrow \sqrt{\left<\vec{m}^2\right>} \ ,
\label{linearize.rho}
\end{equation}
which would be exact in the large-$n$ limit.
For convenience we set the external field in the $(1,1,...)$ direction
\begin{equation}
\vec{h}(t) = \frac{h(t)}{\sqrt{n}}\:(1,1,...) \ . \label{h.11}
\end{equation}
Clearly, the scaling functions will not depend on this arbitrary choice.
As in the scalar case we also consider an initial state with a uniform
bias. For simplicity we restrict the bias to be parallel or antiparallel
to the external field:
\begin{equation}
\left<\vec{m}({\bf r},0)\right> = m_0\:(1,1,...) \ . \label{m0.11}
\end{equation}
{}From (\ref{vmh.eq.3}), (\ref{h.11}) and (\ref{m0.11}) it follows that the
different components of $\vec{m}$ have the same expectation values at all
times.
Hence $\left<\vec{m}\right>=\left<m\right>(1,1,...)$ and
$\left<\vec{m}^2\right>= n\left<m^2\right>$, where $m$ is any component of
$\vec{m}$. Inserting (\ref{linearize.rho}) and (\ref{h.11}) in
(\ref{vmh.eq.3}), we obtain the self-consistent linear equation for the
components of $\vec{m}({\bf r},t)$:
\begin{equation}
\partial m/\partial t \ = \ \nabla^2m + a(t)\,m + c(t)\,h(t) \label{vmh.eq.4}
\end{equation}
\begin{eqnarray}
a(t) & = & 1 - \left<(\nabla m)^2\right> \ , \nonumber \\
c(t) & = & \sqrt{\left<m^2\right>} \ = \ \sqrt{S_0(t)+\left<m\right>^2} \ ,
\label{ct.1}
\end{eqnarray}
which is the vector field version of (\ref{hm.eq.3}).

The calculation of the expectation values of $m$ proceeds as in the previous
subsection, the only difference is that $c(t)$, or equivalently $M(t)$ or
$\eta(t)$, has to be determined self-consistently whenever $h\neq 0$.
As usual we adopt a gaussian distribution with mean $m_0$ and second cumulant
(\ref{mm0}) for the initial condition of each component of $\vec{m}$.
Fourier transforming (\ref{vmh.eq.4}), and noting that $b(t)$ (defined by
$\dot{b}(t)=a(t)$) is still given by (\ref{bt.2}) for $t>t_0$, yields the same
solution (\ref{hmk.eq.2}) for $m_{{\bf k}}(t)$, but where $\eta(t)$ is now
given by
\begin{equation}
\eta(t) = \int_{t_0}^t dt'\,h(t')\,c(t')\,
          \left(\frac{t_0}{t'}\right)^{\frac{d+2}{4}} \ , \label{eta.n}
\end{equation}
rather than (\ref{eta}).
This leads to the same expressions (\ref{M0}), (\ref{SA.C0}) and (\ref{SA.S0})
for
the single-component averages $\left<m\right>$, $\left<m(1)m(2)\right>_c$ and
$\left<m^2\right>_c$.
Hence the normalized correlator $\gamma(1,2)$ is again given by
(\ref{SA.gammac.t1t2}),
and the dimensionless average $M(t)$ is given by (\ref{SA.M}) but with
$\eta(t)$ obeying (\ref{eta.n}). Using (\ref{ct.1}), (\ref{M}) and
(\ref{SA.S0}) we may rewrite the driving force as:
$c(t)h(t)=\sqrt{4t/d}\,\sqrt{1+2M^2(t)}\,h(t)$.
Inserting this in (\ref{eta.n}), and then (\ref{eta.n}) in (\ref{SA.M}),
gives the explicit equation for $M(t)$:
\begin{equation}
\sqrt{2}M(t) = \left(\frac{t}{t_0}\right)^{d/4} \left[m'_0 +
\int^t_{t_0}dt'\,h(t') \left(\frac{t_0}{t'}\right)^{d/4}\sqrt{1+2M^2(t')}
\right]  \ , \label{M.eq.1}
\end{equation}
with $m'_0=\sqrt{d/4}\,m_0/\sqrt{t_0}$. For $h\neq 0$ this is a self-consistent
equation which has to be solved numerically. It yields a much faster growth for
$M(t)$ than in the scalar case (eq.\ (\ref{SA.M})), where the integral in
(\ref{M.eq.1}) is just (\ref{eta}), i.e.\ $c(t)=1$.
Alternatively, we could write equations for $c(t)$ or $\eta(t)$, but
$M(t)$ is the relevant quantity for the scaling functions.
Equation (\ref{M.eq.1}) may be recast in the differential form
\begin{equation}
\frac{d}{dt}\left(\frac{\sqrt{2}M}{t^{d/4}}\right) = h(t)\,
\sqrt{ \frac{1}{t^{d/2}} + \left(\frac{\sqrt{2}M}{t^{d/4}}\right)^2 } \ ,
\label{M.eq.2}
\end{equation}
with initial condition at $t=t_0$
\begin{equation}
\left(\frac{\sqrt{2}M}{t^{d/4}}\right) = \frac{m'_0}{t_0^{d/4}}
\ \ \ , \ \ \ t=t_0 \ . \label{M.IC.2}
\end{equation}
This is equivalent to the equation obtained by BK \cite{BK:92} in the large-$n$
treatment of the dynamics of $\vec{\phi}$.

As suggested by (\ref{M.eq.2}), we find that $M$ is a scaling function of $t$
and $h(t)$. To show this it is convenient to separate the cases $h={\rm const}$
and $h=h(t)$.
First we consider a time-independent external field $h$. Without loss of
generality we set $h>0$, while allowing the initial bias to be $m_0>0$ or
$m_0<0$. Defining the scaling function and variables
\begin{eqnarray}
N(y,z) & = & \sqrt{2}M(t)/y^{d/4} \ , \label{Nh} \\
     y & = & ht                   \ , \label{y} \\
     z & = & m'_0/(ht_0)^{d/4} \ = \ m'_0/y_0^{d/4}  \ , \label{z}
\end{eqnarray}
equation (\ref{M.eq.2}) then reads
\begin{equation}
\frac{\partial N}{\partial y} = \sqrt{\frac{1}{y^{d/2}} + N^2} \ ,
\label{Nh.eq}
\end{equation}
with initial condition at $y_0=ht_0$
\begin{equation}
N(y_0,z) = z \ , \label{Nh.IC}
\end{equation}
where $y_0$ is a dimensionless measure of the magnetic field.
Note that $N$ only depends on the initial bias through the initial condition.
The solutions of (\ref{Nh.eq}) for large-$y$, and for small-$y$ and $z$
(with logarithms for $d=4$), are
\begin{eqnarray}
N(y,z) & \simeq & B\,\exp(y) \ \ , \ \ y\gg 1 \label{Nh.large-y} \\
N(y,z) & \simeq & A\,y^{(4-d)/4} + \left[z-A\,y_0^{(4-d)/4}\right] \ \ , \ \
y\ll 1 \ , \ d\neq 4 \label{Nh.small-y}
\end{eqnarray}
where $A=4/(4-d)$ and $B$ is a constant.
Using $\sqrt{2}M=N\,y^{d/4}$ gives the corresponding behaviour of $M(t)$.
Note that for a negative bias $N$ starts negative, but since from (\ref{Nh.eq})
its growth is always positive, $N$ becomes large and positive for large-$y$
($B>0$). This can also be seen from (\ref{M.eq.1}): if $M$ starts large, there
is a feedback through the time integral which boosts the effect of the $h$
field. Therefore, contrary to the scalar case (sec.\ \ref{off.1}) the external
field will always reverse the magnetization, however large $|z|$ may be.
The inversion time cannot be estimated, however, without solving the full
equation (\ref{Nh.eq}). For example, even if we set $M(t)=0$ in (\ref{M.eq.1})
we cannot estimate the continuated effect of $M(t')$ from $t_0$ to $t$.
But if the initial bias is very small, i.e.\ if $|z|$ is small, $N$ will be
small from the initial time to the inversion time.
Then we can set $N=0$ in (\ref{Nh.small-y}), giving (cf.\ (\ref{tinv.1}))
$t_{inv}/t_0 = \exp\left(|m_0|/|ht_0|\right)$ for $d=4$ and
\begin{eqnarray}
\!\!\!\!
y_{inv}^{\frac{4-d}{4}} = y_0^{\frac{4-d}{4}} + |z|/A \ \ , \ \ {\rm or} \ \
\left(t_{inv}/t_0\right)^{\frac{4-d}{4}} = 1 + |m'_0|/|ht_0|A
\ \ , \ \ {\rm for} \ \ d\neq 4 \ . \label{tinv.n}
\end{eqnarray}
The late-time exponential growth of $M(t)$, which will lead to a very rapid
saturation of $\left<\vec{\phi}\right>$, can be understood from the non-linear
equation of motion (\ref{vmh.eq.3}): keeping the important terms at late-times
and writing $\vec{m}=\rho\hat{m}$, yields $\dot{\hat{m}}=\vec{h}_\perp$ and
$\dot{\rho}=\rho\vec{h}_\parallel$.
Hence $\rho\sim\exp(h_\parallel t)$. So, despite the large-$n$ treatment of
$\vec{m}$ we self-consistently obtain the correct asymptotics.
The scaling regime (characterized by the scale $L(t)\gg\zeta$) comprises both
the `initial growth' when $M\ll 1$ (eq.\ (\ref{Nh.small-y})), and the
intermediate stage prior to saturation (when $M\gg 1$). So we want to solve
(\ref{Nh.eq}) (numerically) to obtain the full solution for $M(t)$.
The above scaling form, not only simplifies the equations but also provides a
more transparent picture of the combined role of the variables $t$, $h$ and
$m_0$. Namely, $1/h$ and $(ht_0)^{d/4}\sqrt{t_0}$ are the characteristic scales
of time and bias.

We now consider a time-dependent external field $h(t)$.
For $h(t)\geq 0$ and $h(t)=0$ at isolated points only, we may define the
following one-to-one mapping between the scaling variable and the time $t$:
\begin{equation}
y(t) = \int_0^t dt'h(t') \ . \label{yt}
\end{equation}
Let $t(y)$ be the inverse function of $y(t)$.
By analogy with the previous case we introduce $h_0$, a characteristic
amplitude of $h(t)$. Then, defining the scaling function
\begin{eqnarray}
N(y(t),z) & = & \sqrt{2}M(t)/(h_0t)^{d/4} \ , \label{Nht} \\
        z & = & m'_0/(h_0t_0)^{d/4}    \ , \label{zt}
\end{eqnarray}
and noting that $\partial N/\partial t=h(t)(\partial N/\partial y(t))$,
equation (\ref{M.eq.2}) becomes
\begin{equation}
\frac{\partial N}{\partial y} = \sqrt{(h_0t(y))^{-d/2} + N^2} \ ,
\label{Nht.eq}
\end{equation}
with initial condition (\ref{Nh.IC}) at $y(t)=y_0=h_0t_0$.
The solution of (\ref{Nht.eq}) for $y$ and $z$ both small is
\begin{equation}
N(y,z) \ \simeq \ z + \int_{h_0t_0}^{y(t)} dy'\left(h_0t(y')\right)^{-d/4}
\ \ , \ \ y\ll 1 \label{Nht.small-y}
\end{equation}
while the large-$y$ solution is (\ref{Nh.large-y}), but with $y$ given by
(\ref{yt}). When $h(t)=const$ this formalism reduces to the previous one.
The procedure to relate the scaling variables $y_2(t)$ and $y_1(t)$, concerning
two different fields $h_2(t)$ and $h_1(t)$, is to replace $h_2(t')$ by
$h_2(t(y_1(t')))$ in the time integral defining $y_2(t)$.
In general the function $y(t)$ will not be invertible, however, so the scaling
form is of limited practical use in this case. For example, the integral of
$h(t)=h_0\exp(-\omega t)$ is easily inverted, but the integral of
$h(t)=h_0(1-\sin(\omega t))$ has to be inverted numerically. In practice, of
course, one can always solve equation (\ref{M.eq.2}) for each given field
$h(t)$.
{}From a theoretical point of view, however, the scaling variable $y(t)$ is
interesting: it shows that all that matters at a given time, is the sum of the
field at all previous times.

To evaluate the expectation values for the order parameter $\vec{\phi}$ in the
scaling regime, we make the usual replacement $\vec{\phi}(\vec{m})\to\hat{m}$.
Given the gaussian initial conditions and the linearity of (\ref{vmh.eq.4})
(following from taking the large-$n$ limit), the components of
$\vec{m}({\bf r},t)$ are independent gaussian variables with non-zero mean.
Evaluating the gaussian averages we obtain
$n$-dependent scaling functions with arguments $M(t)$, $\gamma(1,2)$ and/or
$S_0(t)$, calculated above in the large-$n$ limit.
We recall that, as far as the determination of the $\vec{m}$ moments and of the
scaling functions is concerned, the leading order of the systematic approach,
i.e.\ the large-$n$ treatment of the $\vec{m}$ equation, is equivalent to
assuming that $\vec{m}({\bf r},t)$ is a gaussian field.

\bigskip

\noindent
{\bf Magnetization} \\
\noindent
Evaluating the gaussian average of the field $\vec{\phi}$, we obtain
(see the Appendix)
\begin{eqnarray}
\left<\vec{\phi}(t)\right> & = & \left<\frac{\vec{m}}{|\vec{m}|}\right>
\nonumber \\
              & = & \vec{M}(t)\,\frac{2}{\sqrt{\pi}} \int_0^1 ds\,
(1-s^2)^{\frac{n-1}{2}} \exp\left(-\vec{M}^2(t)s^2\right) \ , \label{phi.n}
\end{eqnarray}
where $\vec{M}(t)=M(t)(1,1,...)$ follows from (\ref{h.11})-(\ref{m0.11}), and
$M(t)$ is the solution of equation (\ref{M.eq.2}). This result reduces to the
error function form (\ref{phi}) when $n=1$.
In the early scaling regime ($M^2\ll 1$) we can expand the exponential in
(\ref{phi.n}). Then, from (\ref{M.eq.1}), or using (\ref{Nht.small-y}) and
(\ref{Nht}), we obtain
\begin{eqnarray}
|\left<\vec{\phi}(t)\right>| & \simeq &
\frac{B\left(1/2,(n+1)/2\right)}{\sqrt{\pi}}\,|\vec{M}(t)| \ \ \ , \ \ \
|\vec{M}(t)|\ll 1 \nonumber \\
& = & \frac{B\left(1/2,(n+1)/2\right)\sqrt{n}}{\sqrt{\pi}}\,\left[
\left(\frac{t}{t_0}\right)^{d/4}m'_0 +
\int^t_{t_0}dt'h(t')\left(\frac{t}{t'}\right)^{d/4} \right]
\label{phi.n.small-M}
\end{eqnarray}
in close agreement with the predictions from the large-$n$ treatment of
$\vec{\phi}$ \cite{BK:92}.
In the saturation limit ($M^2\gg 1$) the integral in (\ref{phi.n}) is dominated
by small values of $s$ and we can expand the binomial. Using (\ref{Nh.large-y})
and (\ref{Nht}) for $h\neq 0$, and (\ref{M.eq.1}) for $h=0$, this gives
\begin{eqnarray}
|\left<\vec{\phi}(t)\right>| & = & 1 - \frac{n-1}{4\vec{M}^2} + O(1/\vec{M}^4)
\ \ \ , \ \ \ |\vec{M}(t)|\gg 1 \label{phi.n.large-M} \\
& \simeq & 1 - \frac{n-1}{2n} \,
\frac{\exp\left(-2\int^t_0dt'h(t')\right)}{(t/t_0)^{d/2}} \ \ , \ \ h\neq 0
\label{phi.n.large-Mh} \\
& \simeq & 1 - \frac{n-1}{2n} \,
\frac{1}{m_0^{'2}(t/t_0)^{d/2}} \ \ , \ \ h=0 \ . \label{phi.n.large-Mm0}
\end{eqnarray}
Hence, the magnetization saturates exponentially fast in the presence of an
external field, and as a power-law when there is only an initial bias. Again,
this agrees with the large-$n$ limit of $\vec{\phi}$ \cite{BK:92,PFT:87}.
As expected, the magnetization induced by the external field always dominates
over the initial bias at late-times.
In other words, the system loses the memory of its initial condition.
This differs from the scalar case, where both $h$ and $m_0$ contribute to the
exponential behaviour (cf.\ (\ref{phi.large-M})).
The magnetization curves for the full time-range are shown in Figure 2 for
different values of $n$ and $h$.

Although the transverse magnetization is zero by symmetry, the coefficient
$(n-1)$ in (\ref{phi.n.large-M}) suggests that the leading correction to
saturation is due to the part of $\vec{\phi}$ transverse to $\vec{h}$.
This is easy to understand:
recall that $\vec{\phi}$ quickly saturates in length in any direction, hence
the time increase in the net component of $\vec{\phi}$ along $\vec{h}$ must
come from rotation.

\bigskip

\noindent
{\bf Pair correlations} \\
\noindent
The gaussian calculation of the two-time pair correlation function of
$\vec{\phi}$ is relatively straightforward, though the algebra is rather
long (see the Appendix). The simplest way we found to express the result is
the following:
\begin{eqnarray}
\!\!\!\!\!\!\!\!\! & \!\!\!\!\!\!\!\!\! & \!\!\!\!\!
C(1,2) \ = \ \left<\hat{m}(1)\!\cdot\!\hat{m}(2)\right>  \nonumber \\
\!\!\!\!\!\!\!\!\! & \!\!\!\!\!\!\!\!\! & \!\!\!\!\!
= \frac{4}{\pi}  \int_0^1du \int_0^1dv
\left[ \frac{n}{2}\gamma + \vec{M}(1)\!\cdot\!\vec{M}(2) -
       \frac{\alpha(u,v)}{\beta(u,v)}\gamma \right]
\frac{ \left((1\!-\!u^2)(1\!-\!v^2)\right)^{\frac{n-1}{2}}}
     {\beta(u,v)^{\frac{n+2}{2}} }
\exp\left\{ -\frac{\alpha(u,v)}{\beta(u,v)} \right\}, \nonumber \\
\label{SAh.Cn}
\end{eqnarray}
\begin{eqnarray}
\alpha(u,v)
& = & \vec{M}^2(1)u^2+\vec{M}^2(2)v^2 -
2\gamma(1,2)\vec{M}(1)\!\cdot\!\vec{M}(2)u^2v^2  \nonumber \\
\beta(u,v) & = & 1 - \gamma^2(1,2)\, u^2v^2 \ . \label{SAh.Cn.aux}
\end{eqnarray}
$\vec{M}(i)$ is the same as in (\ref{phi.n}), and $\gamma(1,2)$ is given by
(\ref{SA.gammac.t1t2}).
The equal-time correlation function $C(x,M)$ is show in Figure 3 for
different values of $n$ and constant field $h$ (the same values as in
Figure 2).

An especially convenient algorithm for numerical integration is obtained with
the substitutions $u=\cos(a)$ and $v=\cos(b)$: this yields a factor
$(\sin(a)\sin(b))^n$ in the integrand making its singularity for $\gamma=1$
(equal-times) explicitly integrable.
The scaling function (\ref{SAh.Cn}) generalizes the usual BPT scaling function
to an off-critical system, and indeed it reduces to (\ref{BPT}) when $M=0$.
It can be checked that any choice of axis, different from (\ref{h.11}),
produces the same result (\ref{SAh.Cn}).
The transverse and longitudinal parts of $C(1,2)$ (relative to $\vec{h}$) can
be read off from the big bracket in the integrand: the transverse
contribution arises from the term $(n-1)\gamma/2$, while the longitudinal one
arises from $(1/2-\alpha/\beta)\gamma + \vec{M}(1)\!\cdot\!\vec{M}(2)$.
When $n=1$ (\ref{SAh.Cn}) should reduce to (\ref{SAh.C})
(with $\left<\phi\right>$ given by (\ref{phi})).
To show this one must express (\ref{SAh.Cn}) as a single integral using
appropriate variable changes. Although we have not been able to give an
analytical proof, we have verified the equivalence to great numerical accuracy.
Another check on (\ref{SAh.Cn}) is the large-$n$ limit, which gives
$\left<\vec{m}(1)\!\cdot\!\vec{m}(2)\right>/
(\left<\vec{m}^2(1)\right>\left<\vec{m}^2(2)\right>)^{1/2}$, as expected.

When $\gamma\to 0$ with fixed $M(1)$ and $M(2)$, i.e.\ $r\to\infty$
with fixed times, the correlations decay as
(compare (\ref{SAh.C.large-x}) for $n=1$)
\begin{equation}
C(1,2) = \left<\vec{\phi}(1)\right>\!\cdot\!\left<\vec{\phi}(2)\right> +
a(t_1,t_2)\,\gamma + O(\gamma^2) \ , \label{SAh.Cn.large-r}
\end{equation}
where $a(t_1,t_2)$ is a known function of $\vec{M}(1)$ and $\vec{M}(2)$.
It also interesting to look at the decay of the autocorrelation function
$A(t_1,t_2)=C(0,t_1,t_2)$ when $t_2\gg t_1=t_0$, $h=0$ and $m_0\neq 0$.
In this case (\ref{M.eq.1}) and (\ref{SA.gammac.t1t2}) give:
$M(1)\sim m_0'$, $M(2)\sim M(1)/\gamma$ and $\gamma\sim (t_1/t_2)^{d/4}$.
A careful expansion of (\ref{SAh.Cn}) then yields
\begin{equation}
A(t_1,t_2) \ = \ |\left<\vec{\phi}(t_1)\right>|\:\left[1 +
\frac{(n-1)B}{m_0^{,2}}\left(\frac{t_1}{t_2}\right)^{d/2} +
O(\gamma^4) \right] \ \ , \ \ t_2\gg t_1 \ , \label{SA.A.decay}
\end{equation}
where $B=(2^{(d+2)/2}-1)/2$. As expected, the saturation limit of the
autocorrelations is the initial bias in the field $\vec{\phi}$.
The coefficient $(n-1)$ indicates a contribution from rotation, as in
(\ref{phi.n.large-M}).

\bigskip

\noindent
{\bf Density of defects} \\
\noindent
Finally, we evaluate the average density of defect core volume per unit
volume of the system, $\rho_{def}(t)$. The gaussian calculation follows
the same steps as in the previous section for a scalar system. This gives
\begin{eqnarray}
\rho_{def}(t) & = &
\left<\delta(\vec{m})\right> \label{aver.rho.vm} \ = \
\left(P(0)\right)^n \nonumber \\
 & = & \left(\frac{d}{8\pi}\right)^{n/2}\:
\frac{\exp\left(-\vec{M}^2(t)\right)}{t^{n/2}} \ , \label{SAh.rho.vm}
\end{eqnarray}
which generalizes (\ref{SAh.rho.m}). $P(m)$ is the one-point distribution
for each component of $\vec{m}$, given by (\ref{Pm}).
For $m_0=h=0$ (\ref{SAh.rho.vm}) gives the usual scaling form
$\rho_{def}\sim L^{d-n}/L^d \sim 1/L^n$.

\subsection{Comparison with experiment: $2d \ XY$ model with bias}
\label{experiment}

For systems with $d=n=2$ the defects are vortices and anti-vortices
interacting strongly below the transition temperature
(known as `Kosterlitz-Thouless' transition temperature; see e.g.\
\cite{G:92}).
The energy of a vortex pair is of order $\ln(L/\zeta)$, with $\zeta$ the
vortex core size and $L$ the vortex pair separation.
If scaling holds, then the defect density $\rho_{def}(t)$ should scale as
$L^{-2}(t)$ (eq.\ (\ref{SAh.rho.vm})). The product of these two results
then gives the energy density, $\epsilon \sim \ln(L/\zeta)/L^2$.

The planar $XY$ model is of particular interest since it appears to be a
special case: there are no conclusive predictions or definitive measurements
regarding the asymptotic growth in this systems, but there are indications
that the growth is anomalous.
Yurke et al.\ \cite{YPKH:93} have suggested a slower growth law
$L\sim (t/\ln t)^{1/2}$ for non-conserved systems, and indeed it has proven
difficult to reach the scaling regime through numerical simulations
(see e.g.\ \cite{PGY:94,YPKH:93} and references therein).
More recently, Bray and Rutenberg \cite{BR:94} concluded that the energy
dissipation due to the ordering process occurs significantly on all scales
between $\zeta$ and the inter-vortex spacing $L(t)$, suggesting that there may
not be a single characteristic length scale in the system, i.e.\ that
the scaling hypothesis may not hold in this case. On the assumption that
scaling {\em does} hold however, the $(t/\ln t)^{1/2}$ growth is recovered
\cite{BR95}.
Numerical simulations \cite{BB:94} also provide evidence that the scale
$L(t)$ required to collapse the data for the pair correlation function,
and the typical intervortex spacing $\rho(t)^{-1/2}$, are not simply
proportional to each other.

The systematic approach cannot account for these possible logarithmic effects
since, as a consequence of treating the dynamics of the field $\vec{m}$
in the large-$n$ limit, a $t^{1/2}$ growth is predicted for all $n$ and $d$.
For the same reason, it also gives the incorrect value $d/2$ for the exponent
$\lambda$ (eq.\ (\ref{SA.lambda})).
It is still interesting, however, to make a qualitative comparison between the
theory and experimental or simulation data, namely to confront the effect of
the initial bias in the results.

Pargellis, Green and Yurke (PGY) \cite{PGY:94} have devised an experimental
system exhibiting planar $XY$-model behaviour.
The system consists of a nematic liquid crystal material placed between two
plates. A suitable choice of the temperatures at the two plates creates a
nematic-isotropic interface (parallel to the plates) near the centre of
the cell, where the nematic director has a fixed angle with the normal axis.
Hence the projection of the director onto the interface provides the $xy$
degree of freedom. In addition, a normal alignment is imposed at the colder
plate, creating a splay of the director from this plate to the interface.
With these boundary conditions the symmetry $\vec{\phi}\to -\vec{\phi}$ is
effectively broken (i.e.\ there are no defects with charge $1/2$) and at the
interface the defects look like vortices.
After inducing a thermal quench in this system, PGY measured the
density of defects $\rho(t)=\rho_{def}(t)$, the autocorrelation function
$A(t,t_0)$ and the magnetization $\left<\vec{\phi}(t)\right>$.
To confirm that the system is governed by the $XY$-model dynamics, PGY
performed numerical simulations of this model using as initial condition, not
a random field configuration, but rather one obtained from experiment.
Although they found good agreement between the results, these did not exhibit
the expected scaling behaviour $\rho(t)\sim t^{-1}$ and
$A(t,t_0)\sim t^{-(d-\lambda)/2}\sim t^{-0.585}$ (where the latter follows from
Mazenko's theory \cite{BH:92,LM:92:lambda.sim}), indicating that some aspect of
the initial condition was preventing the system from reaching the scaling
regime.
To study the effect of a non-uniform distribution for the angle $\theta$
between the initial field $\vec{\phi}(t_0)$ and the $x$ axis, PGY performed
simulations using the probability distribution:
\begin{equation}
P(\theta) \ = \ \frac{1}{2\pi}\left[1+A\,\cos(k\theta)\right] \ , \label{Pt}
\end{equation}
where $A$ is the `bias parameter', and $k=1$ corresponds to a dipolar deviation
from uniformity. Higher order distortions ($k\geq 2$) were shown to produce a
late-time behaviour indistinguishable from the one with zero-bias ($A=0$), and
therefore were excluded.
The log-log plots of the simulation data (with different values of $A$) for
$\rho(t)$, $A(t,t_0)$ and
$|\left<\vec{\phi}(t)\right>|/|\left<\vec{\phi}(t_0)\right>|$ are
shown in Figures 4, 6, and 8. The corresponding log-log plots from the
systematic approach (sec.\ \ref{off.n}), obtained using the relation
$M(t_0)=A/\sqrt{2\pi}$ (see below), are shown in Figures 5, 7, and 9.
Taking into account that the scales of time are arbitrary,
there appears to be a good qualitative agreement between the theory and the
simulations, namely in the manner in which the curves change with the bias
parameter $A$.
PGY found that the best fit of the simulations to the experimental data is for
$A=0.07$.
At late times, the simulation data and the theoretical curves for $\rho(t)$
(Figures 4 and 5) drop below the line $t^{-1}$ (the expected decay for $A=0$),
as a consequence of the initial bias. This is what we expect from
(\ref{SAh.rho.vm}) and (\ref{M.eq.1}), which give:
$\rho(t)\sim\exp(-t\,m_0^2)/t$.
{}From the zero-bias autocorrelations (expected to decay as
$t^{-(d-\lambda)/2}$)
PGY measured $(d-\lambda)/2=0.543\pm 0.009$, while the corresponding results
from Mazenko's theory and from the systematic approach are
$(d-\lambda)/2=0.585$ (solid line in Figure 6) and $(d-\lambda)/2=0.5$
(solid line in Figure 7).
For $A\neq 0$, the autocorrelation curves do not decay indefinitely, but have a
saturation limit which is the initial bias
$\phi_0=|\left<\vec{\phi}(t_0)\right>|$.
This is what we expect from (\ref{SA.A.decay}) which gives:
$\ln(A(t,t_0))\simeq\ln(\phi_0)+1/(m_0^2t)^{d/2}$ for $t\gg t_0$.
The solid lines in Figures 8 and 9, with slopes $\lambda=0.83$ and
$\lambda=1$, describe the initial growth of the magnetization
(cf.\ (\ref{phi.n.small-M})).
The fits to the simulation data ($A=0.07$) and to the experimental data,
however, give $\lambda=0.518\pm 0.021$.
The magnetization curves with the larger values of $A$ indicate that the
system is close to saturation, and therefore outside the scaling regime.
This behaviour is described by (\ref{phi.n.large-Mm0}).
Since the experimental and simulation results do not respect the usual scaling
forms in part because of the bias, they are inconclusive about the existence of
logarithmic factors in $L(t)$.

In order to compare the theory with the PGY data, we calculated the
probability distribution of $\theta$ corresponding to a gaussian
distribution of the initial auxiliary field $\vec{m}({\bf r},t_0)$.
Since $P(\theta)$ in (\ref{Pt}) is maximum for $\theta=0$, $\theta$ is the
angle between $\vec{m}=\vec{m}(t_0)$ and the most probable direction
$\vec{m}_0\equiv\left<\vec{m}(t_0)\right>$, i.e.\ $\mu=\cos(\theta)=
\vec{\phi}(t_0)\!\cdot\!\left<\vec{\phi}(t_0)\right>=\hat{m}\!\cdot\!\hat{m}_0$.
Noting that $P(\vec{m})$ is the product of two distributions of the
form (\ref{Pm}), using polar coordinates, exploiting the symmetry of
$\cos(\theta)$ and making the substitution $s=\cos(\theta)$, we obtain
\begin{eqnarray}
P(\mu) & = & \int d\vec{m}\,P(\vec{m})\delta(\mu-\hat{m}\!\cdot\!\hat{m}_0)
\nonumber \\
 & = & \int_0^{\infty}\frac{dm\,m}{2\pi S_0}\,
\exp\left(-\frac{m^2+m_0^2}{2S_0}\right)
2\int_{-1}^{1}\frac{ds}{\sqrt{1-s^2}}\,\exp\left(\frac{m\,m_0\,s}{S_0}\right)\,
\delta(\mu-s) \ . \nonumber
\end{eqnarray}
Performing the integrals and noting that $P(\theta)=P(\mu)|\sin(\theta)|/2$,
gives
\begin{equation}
P(\theta) \ = \ \frac{\exp\left(-M_0^2\right)}{2\pi}\left[ 1 +
M_0\sqrt{\pi}\cos(\theta) \exp\left[M_0^2\cos^2(2\theta)\right]
\left(1+{\rm erf}(M_0\cos(\theta))\right) \right] \ , \label{Ptt}
\end{equation}
where $M_0 = |\vec{M}_0| = m_0/\sqrt{2S_0}$ and $m_0=|\vec{m}_0|$
is the initial bias.
Expanding (\ref{Ptt}) when $M_0$ is small yields, to leading order,
\begin{equation}
P(\theta) \ = \ \frac{1}{2\pi}\left[ 1 + M_0\sqrt{\pi}\cos(\theta) \right] \ .
\label{Pttt}
\end{equation}
Comparing with (\ref{Pt}) gives $M(t_0) = M_0/\sqrt{2} = A/\sqrt{2\pi}$,
which we used to calculate the theoretical plots for given values of $A$.

\section{Conclusions}
\label{end.3}

We have extended the systematic approach (to leading order) to investigate the
asymptotic ordering dynamics of non-conserved quenched systems with an
external field and an initial bias.
An important difference relative to critical quenches is that, as the phase
symmetry is broken the system evolves more rapidly towards final equilibrium.
In particular, as the topological defects disappear more rapidly the scaling
regime, characterized by the length-scale $L(t)\sim\sqrt{t}$, has a more
limited duration.

A key ingredient of the approach is to identify the important effect of the
external field in the ordering process, which is a wall driving force for
scalar systems and bulk rotation for vector systems. This allowed us to make
convenient choices for the potential driving forces ($V_1'(\phi;h)$ and
$\vec{U}(\vec{\phi};\vec{h})$) and to introduce a meaningful auxiliary field.
Treating the auxiliary field dynamics in the large-$d$ (or large-$n$) limit,
we were then able to calculate the magnetization (eqs.\ (\ref{phi}),
(\ref{phi.n})), the pair correlation function (eqs.\ (\ref{SAh.C}),
(\ref{SAh.Cn}), which extend the standard OJK and BPT scaling functions to
off-critical quenches) and the density of defects (eqs.\ (\ref{SAh.rho.m}),
(\ref{SAh.rho.vm})), for scalar and for vector systems.

The magnetization and the pair correlations are shown in Figures 1 to 3,
for particular choices of $n$, $d$, $m_0$ and $h(t)$.
These results have been studied in various limiting cases:
eqs.\ (\ref{phi.large-M})-(\ref{phi.small-M}) for $\left<\phi\right>$,
eqs.\ (\ref{phi.n.small-M})-(\ref{phi.n.large-Mm0})
for $\left<\vec{\phi}\right>$,
eqs.\ (\ref{SAh.C.small-x})-(\ref{SAh.C.large-x}) for $C(1,2)$ with $n=1$ and
eqs.\ (\ref{SAh.Cn.large-r})-(\ref{SA.A.decay}) for $C(1,2)$ with $n>1$.
The respective behaviours have been discussed and are well understood and in
accord with our expectations.
We find, in particular, that the saturation of the magnetization and the decay
of the defect density are exponentially fast.
The initial-growth of the magnetization
(eqs.\ (\ref{phi.large-M})-(\ref{phi.n.small-M})) agrees well with a previous
predictions by BK \cite{BK:92} based both on the large-$n$ solution for
$\vec{\phi}$ and on general scaling arguments.
We also compared our results for a biased $2d$ $XY$-model with numerical
simulation data \cite{PGY:94} (sec.\ \ref{experiment} and Figures 4 to 9),
and found good qualitative agreement.
These simulation data, on the other hand, had been found to agree well (for a
certain value of the initial bias) with data from experiments on
nematic liquid crystals \cite{PGY:94}.

Finally, we recall that the auxiliary field formulation relies on the
existence of stable topological defects, which govern the asymptotic
dynamics of the system. One legitimate question, therefore, is whether these
defects will ever form in the presence of an external field, or a bias, before
the system orders completely.
Clearly, the defects will form, and an intermediate scaling regime will occur
(where $L(t)$ is the typical inter-defect distance) if the field and the
initial bias are not too strong.

\section{Acknowledgments}
We thank A. N. Pargellis et al.\ \cite{PGY:94} for making their
raw data available.
We thank the Isaac Newton Institute, Cambridge, (AB and SP), and the
Physics Departments of the Universities of Illinois and Manchester (SP),
for hospitality. JF thanks JNICT (Portugal) for support.

\section{Appendix: Gaussian Expectation Values}
\label{app.3}

In this Appendix we indicate how to perform the calculation of the
gaussian expectation values which were stated throughout this paper.
The standard results for $h=m_0=0$ can be obtained
as particular cases of the off-critical results.

For a scalar field, $\phi={\rm sgn}(m)$, $m$ is a gaussian variable with
non-zero mean, correlator $\gamma=\left<m(1)m(2)\right>_c/(S_0(1)S_0(2))^{1/2}$
and second moment $S_0=\left<m^2\right>_c$ (eq.\ (\ref{S0c})).
We recall that $M=\left<m\right>/\sqrt{2S_0}$ (eq.\ (\ref{M})).
For vector fields, $\vec{\phi}=\vec{m}/|\vec{m}|$, we conveniently set the
external field $\vec{h}$ and the initial bias $\vec{m}_0$ in the direction
$(1,1,...)$ (eqs.\ (\ref{h.11})-(\ref{m0.11})). Hence, the components of
$\vec{m}$, generally represented by $m$, are independent gaussian variables
with the same moments as above. We recall that
$\vec{M}=\left<\vec{m}\right>/\sqrt{2S_0}=\left<m\right>(1,1,...)$.
We will need the following integral representations
\begin{eqnarray}
{\rm sgn}(m) & = & \frac{1}{i\pi}\: \int_{-\infty}^{\infty}\!
\frac{dz}{z}\,\exp\left(\frac{izm}{\sqrt{2S_0}}\right) \ , \label{sign} \\
\frac{1}{|\vec{m}|} & = & \frac{1}{\sqrt{2\pi S_0}}\: \int_0^{\infty}\!
\frac{dz}{\sqrt{z}}\,\exp\left(-\frac{z\vec{m}^2}{2S_0}\right) \ . \label{mod}
\end{eqnarray}

\subsection{Magnetization}
Using the one-point gaussian distribution (\ref{Pm}), it easy to obtain the
following results:
\begin{eqnarray}
\left<\exp\left(\frac{izm}{\sqrt{2S_0}}\right)\right> & = &
\exp\left(izM-\frac{z^2}{4}\right) \ , \label{e11} \\
\left<\exp\left(-\frac{zm^2}{2S_0}\right)\right> & = &
\frac{1}{\sqrt{z+1}}\:\exp\left(-\frac{zM^2}{z+1}\right) \ , \label{e12} \\
\left<\frac{m}{\sqrt{2S_0}}\,\exp\left(-\frac{zm^2}{2S_0}\right)\right> & = &
\frac{M}{z+1}\:\left<\exp\left(-\frac{zm^2}{2S_0}\right)\right> \ .
\label{e13}
\end{eqnarray}

First we consider a scalar field $\phi$.
Using the representation (\ref{sign}) and the result (\ref{e11}), we can write
the average value of $\phi$ as
\begin{equation}
\left<\phi\right> \ = \ \left<{\rm sgn}(m)\right> \ = \
\frac{1}{i\pi} \int_{-\infty}^{\infty}\!
\frac{dz}{z}\:\exp\left(izM-\frac{z^2}{4}\right) \ . \nonumber \\
\end{equation}
Differentiating with respect to $M$, completing the squares and performing
the integral, gives $d\left<\phi\right>/dM=(2/\sqrt{\pi})\exp(-M^2)$.
Finally, integrating with respect to $M$, with boundary condition
$\left<\phi(M=0)\right>=0$, gives
\begin{equation}
\left<\phi\right> \ = \ \frac{2}{\sqrt{\pi}}\:\int_0^Mdx\exp(-x^2) \ = \
{\rm erf}(M) \ ,  \label{M.1}
\end{equation}
which is the same as (\ref{SA.M}).

Now we consider a vector field $\vec{\phi}$.
Using the representation (\ref{mod}) and the results (\ref{e12})-(\ref{e13}),
and exploiting the rotational symmetry in the $\vec{m}$ space, we can write
the average value of $\vec{\phi}$ as
\begin{eqnarray}
\left<\vec{\phi}\right> & = & \left<\frac{\vec{m}}{|\vec{m}|}\right> \ = \
\frac{(1,1,...)}{\sqrt{\pi}} \int_0^{\infty}\! \frac{dz}{\sqrt{z}}\:
\left<\frac{m}{\sqrt{2S_0}}\exp\left(-\frac{zm^2}{2S_0}\right)\right>
\left<\exp\left(-\frac{zm^2}{2S_0}\right)\right>^{n-1} \nonumber \\
& = & \frac{\vec{M}}{\sqrt{\pi}} \int_0^{\infty}\!
\frac{dz}{\sqrt{z}(z+1)^{\frac{n+2}{2}}}\: \exp\left(-\frac{zM^2}{z+1}n\right)
\ . \nonumber
\end{eqnarray}
Using the substitution $s=\sqrt{z/(z+1)}$, we then obtain
\begin{equation}
\left<\vec{\phi}\right> \ = \ \vec{M}\: \frac{2}{\sqrt{\pi}} \int_0^1\!
\left(1-s^2\right)^{\frac{n-1}{2}} \exp\left(-s^2M^2n\right) \ , \label{M.n}
\end{equation}
which is the same as (\ref{phi.n}), and clearly reduces to (\ref{M.1}) when
$n=1$.

\subsection{Pair correlations}
Using the gaussian joint distribution for the variables
$m(1)$ and $m(2)$ (which may be the same components of $\vec{m}(1)$ and
$\vec{m}(2)$), one can prove the following results,
with the notation $m_1=m(1)$, $S_1=S_0(1)$, $M_1=M(1)$, and
$\delta=1-\gamma^2$,
\begin{eqnarray}
\left<\exp\left(\frac{iz_1m_1}{\sqrt{2S_1}}+\frac{iz_2m_2}{\sqrt{2S_2}}
\right)\right>
& = &
\exp\left(iz_1M_1+iz_2M_2-\frac{z_1^2+z_2^2+2z_1z_2\gamma}{4}\right)
\nonumber \\
\label{en1} \\
\left<\exp\left(-\frac{z_1m_1^2}{2S_1\delta}-\frac{z_2m_2^2}{2S_2\delta}
\right)\right>
& = &
\left(\frac{1-\gamma^2}{B_{12}}\right)^{1/2}\:\exp\left(-\frac{A_{12}}{B_{12}}
\right) \ , \label{en2} \\
\left<\frac{m_1m_2}{2\delta\sqrt{S_1S_2}}
\exp\left(-\frac{z_1m_1^2}{2S_1\delta}-\frac{z_2m_2^2}{2S_2\delta}\right)
\right>
& = &
\frac{1}{B_{12}} \left[\frac{\gamma}{2}+M_1M_2-\gamma\frac{A_{12}}{B_{12}}
\right] \nonumber \\
& & \left<\exp\left(-\frac{z_1m_1^2}{2S_1\delta}-\frac{z_2m_2^2}{2S_2\delta}
\right)\right> \ , \label{en3}
\end{eqnarray}
with the definitions
\begin{eqnarray}
A_{12} & = & M_1^2z_1 + M_2^2z_2 + \Gamma(\gamma)z_1z_2 \nonumber \\
B_{12} & = & (z_1+1)(z_2+1) - \gamma^2 \ , \nonumber
\end{eqnarray}
where $\Gamma(\gamma)$ is given by (\ref{Gamma.y}).

First we consider a scalar field $\phi$.
Using the representation (\ref{sign}) and the result (\ref{en1}), we can write
the correlation function of $\phi$ as
\begin{eqnarray}
C(1,2) & = & \left<{\rm sgn}(m_1){\rm sgn}(m_2)\right> \nonumber \\
& = & - \frac{1}{\pi^2} \int_{-\infty}^{\infty}\!
\frac{dz_1dz_2}{z_1z_2}\:
\exp\left(iz_1M_1+iz_2M_2-\frac{z_1^2+z_2^2+2z_1z_2\gamma}{4}\right) \ .
\nonumber
\end{eqnarray}
Differentiating with respect to $\gamma$, completing the squares and
performing the integrals, gives
$\partial C(1,2)/\partial\gamma =
(2/\pi)\exp\left(-\Gamma(\gamma)\right)/\sqrt{1-\gamma^2}$.
Finally, integrating with respect to $\gamma$, with boundary condition
$C(\gamma=0)=\left<\phi(1)\right>\left<\phi(2)\right>$, gives
\begin{equation}
C(1,2) \ = \ \frac{2}{\pi}\: \int_0^{\gamma}
\frac{dy}{\sqrt{1-y^2}}\:\exp\left(-\Gamma(y)\right) \ + \
\left<\phi(1)\right>\!\left<\phi(2)\right> \ , \label{C.1}
\end{equation}
which is the same as (\ref{SAh.C}).
For critical quenches, $\Gamma=\left<\phi\right>=0$, and (\ref{C.1}) reduces
to the OJK scaling function $(2/\pi) \sin^{-1}\gamma$.

Now we consider a vector field $\vec{\phi}$.
Using the representation (\ref{mod}) (with the substitution
$z\to z/(1-\gamma^2)$) and the results (\ref{en2})-(\ref{en3}), and
exploiting the rotational symmetry in the $\vec{m}$ space, we can write
the correlation function of $\vec{\phi}$ as
\begin{eqnarray}
C(1,2) & = & \left<\frac{\vec{m}_1}{|\vec{m}_1|}\!\cdot\!
\frac{\vec{m}_2}{|\vec{m}_2|}\right> \nonumber \\
& = & \frac{n}{\pi} \int_0^{\infty}\! \frac{dz_1dz_2}{\sqrt{z_1z_2}}\:
\left<\frac{m_1m_2}{2\delta\sqrt{S_1S_2}}
\exp\left(-\frac{z_1m_1^2}{2S_1\delta}-\frac{z_2m_2^2}
{2S_2\delta}\right)\right>
\nonumber \\
&   & \left<\exp\left(-\frac{z_1m_1^2}{2S_1\delta}-
                           \frac{z_2m_2^2}{2S_2\delta}\right)\right>^{n-1}
\nonumber \\
& = & \frac{1}{\pi\delta} \int_0^{\infty}\! \frac{dz_1dz_2}{\sqrt{z_1z_2}}\:
\left[\frac{n\gamma}{2}+\vec{M}_1\!\cdot\!\vec{M}_2-n\gamma\frac{A_{12}}
{B_{12}}\right]\: \left(\frac{1-\gamma^2}{B_{12}}\right)^{\frac{n+2}{2}}
\exp\left(-n\frac{A_{12}}{B_{12}}\right) \ . \nonumber
\end{eqnarray}
Using the substitutions $u=z_1/(z_1+1-\gamma^2)$ and $u=z_2/(z_2+1-\gamma^2)$,
then gives
\begin{eqnarray}
C(1,2) = \frac{4}{\pi}  \int_0^1dudv
\left[ \frac{n}{2}\gamma + \vec{M}_1\!\cdot\!\vec{M}_2 -
       \gamma\frac{\alpha_{12}}{\beta_{12}} \right]\!
\frac{ \left((1\!-\!u^2)(1\!-\!v^2)\right)^{\frac{n-1}{2}}}
     {\beta_{12}^{\frac{n+2}{2}} }
\exp\left\{ -\frac{\alpha_{12}}{\beta_{12}} \right\} \label{C.n}
\end{eqnarray}
which is the same as (\ref{SAh.Cn}), with $\alpha_{12}$ and $\beta_{12}$
given by (\ref{SAh.Cn.aux}).
For critical quenches, $\vec{M}_1=\vec{M}_2=0$, and (\ref{C.n}) reduces to the
BPT function (\ref{BPT}).

\newpage
\singlespacing

\newpage
\doublespacing

\begin{large}
\noindent{\bf Figure Captions}
\end{large}

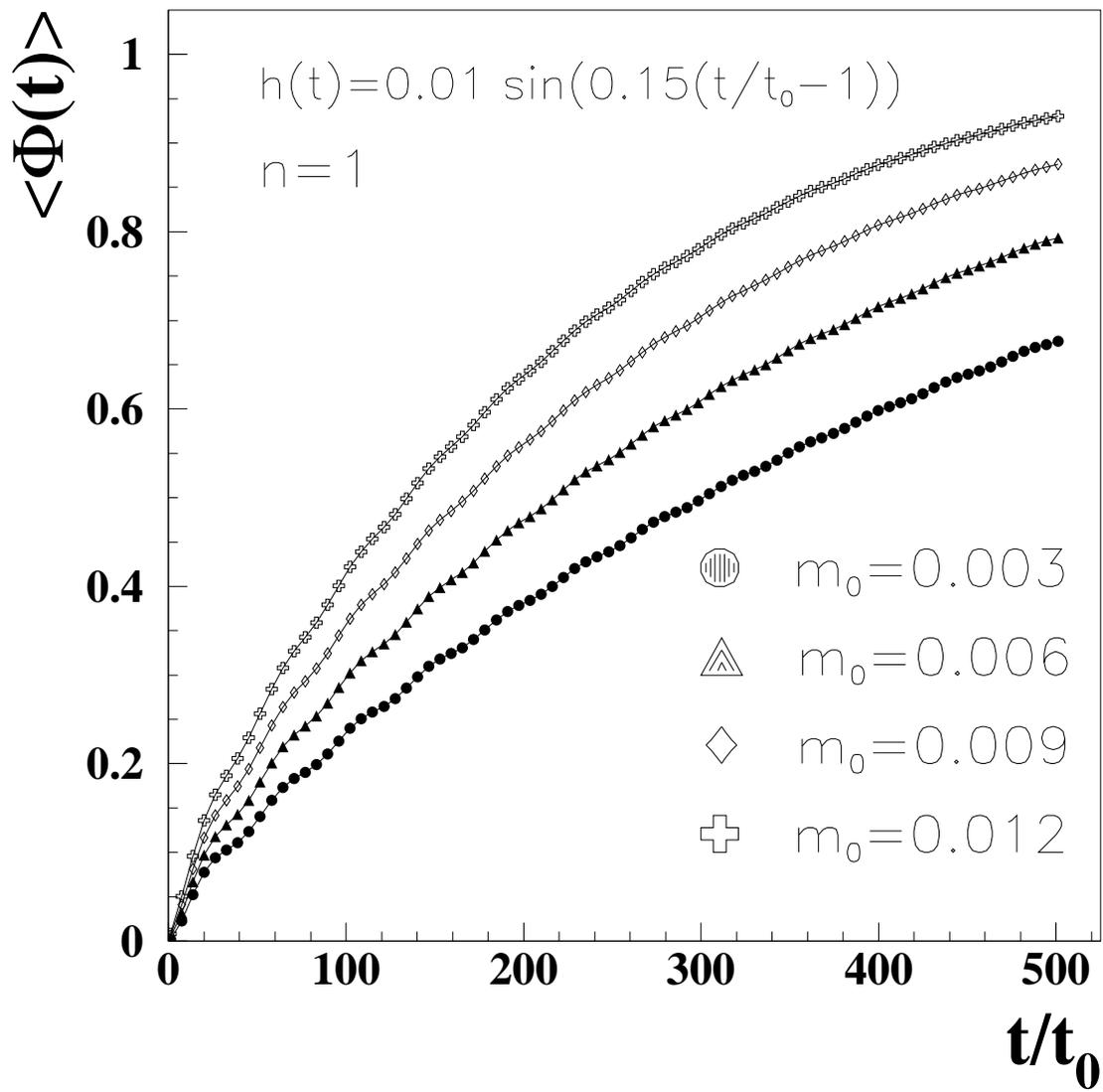
\noindent\underline{Figure 1.} Magnetization for $d=3$ and $n=1$, with a
sinusoidal external field $h(t)=0.01\sin(0.15(t/t_0-1))$ and an initial
bias from $m_0=0.003$ to $m_0=0.012$. Note that $\left<\phi(t_0)\right> >0$.

\medskip

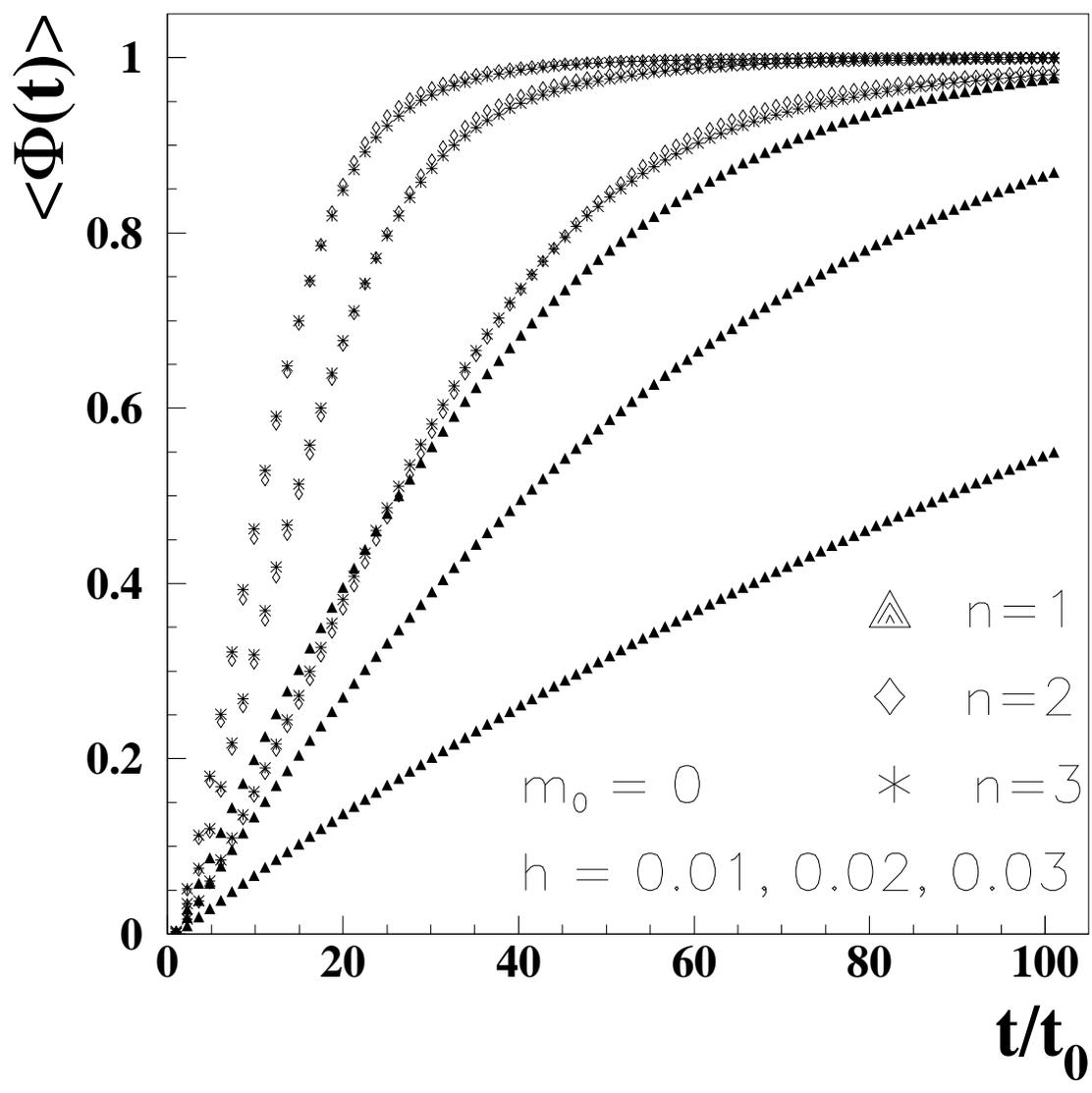
\noindent\underline{Figure 2.} Magnetization for $d=3$ and $n=1,2,3$,
with a constant external field $h=0.01,0.02,0.03$ (bottom to top) and
initial bias $m_0=0$.

\medskip

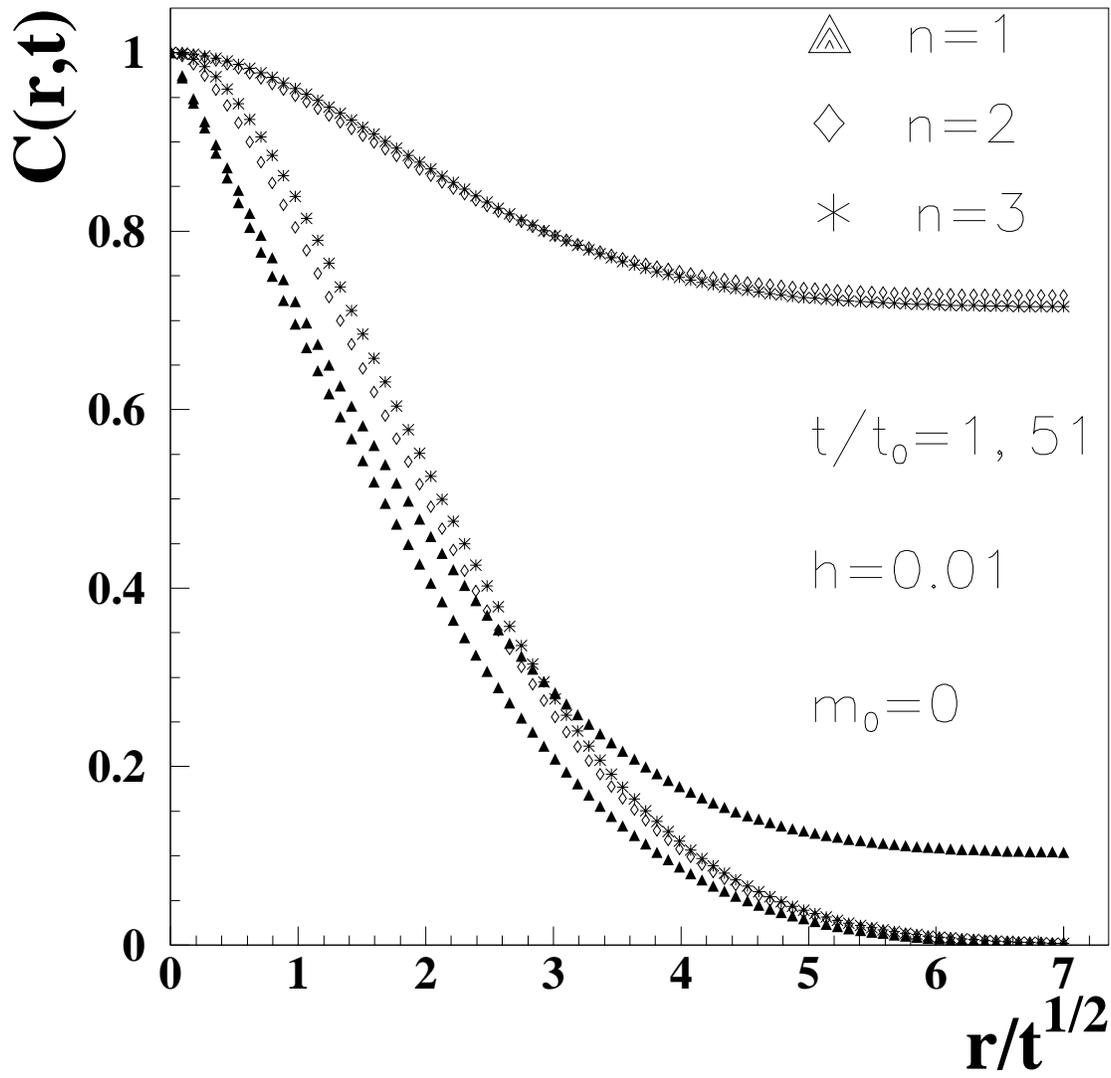
\noindent\underline{Figure 3.} Equal-time pair correlation function for
$d=3$, $n=1,2,3$ and at times $t/t_0=1,51$, with a constant external
field $h=0.01$ and initial bias $m_0=0$. The behaviour at large-$r$ is
described by (\ref{SAh.Cn.large-r}).

\medskip

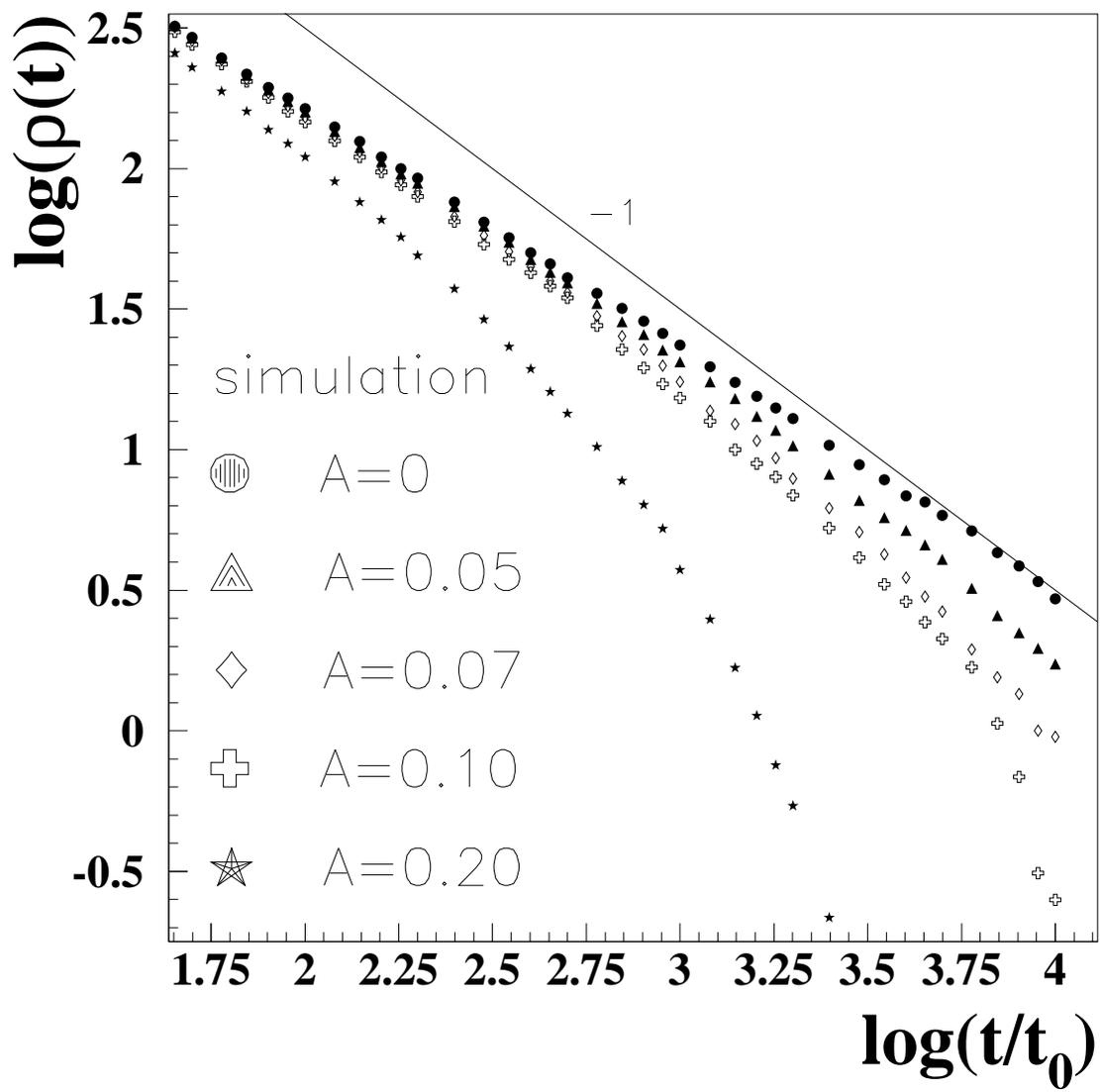
\noindent\underline{Figure 4.} Density of defects for the $2d$ $XY$ model.
Simulation data from Pargellis et al.
There is no applied field and $A$ is the bias parameter (eq.\ (\ref{Pt})).
The solid line with slope $-1$ is the expected asymptotic behaviour.

\medskip

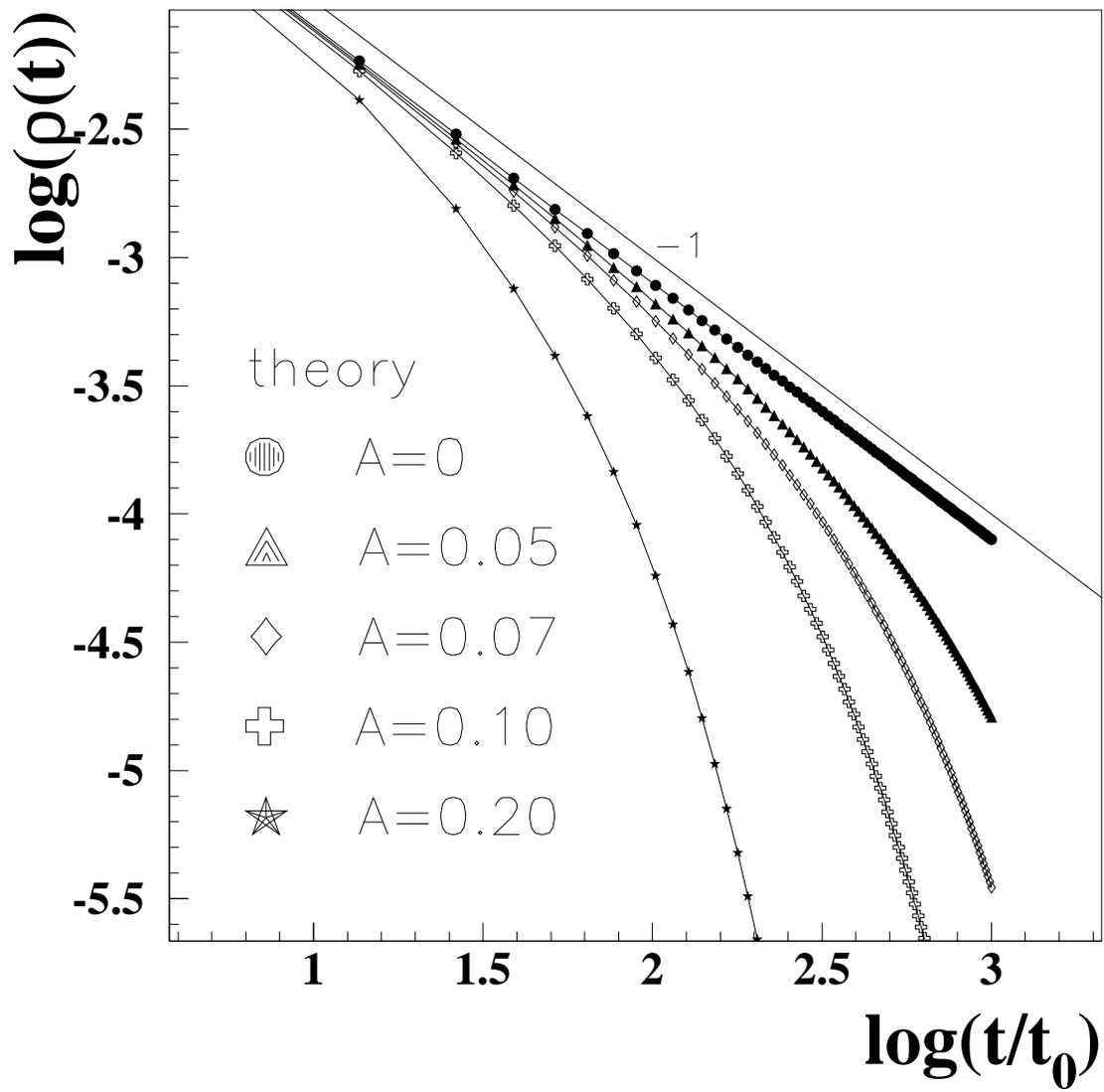
\noindent\underline{Figure 5.} Density of defects for the $2d$ $XY$ model,
as predicted by the systematic approach. There is no applied field, and
the initial bias is given by $M(t_0)=A/(2\pi)^{1/2}$.

\medskip

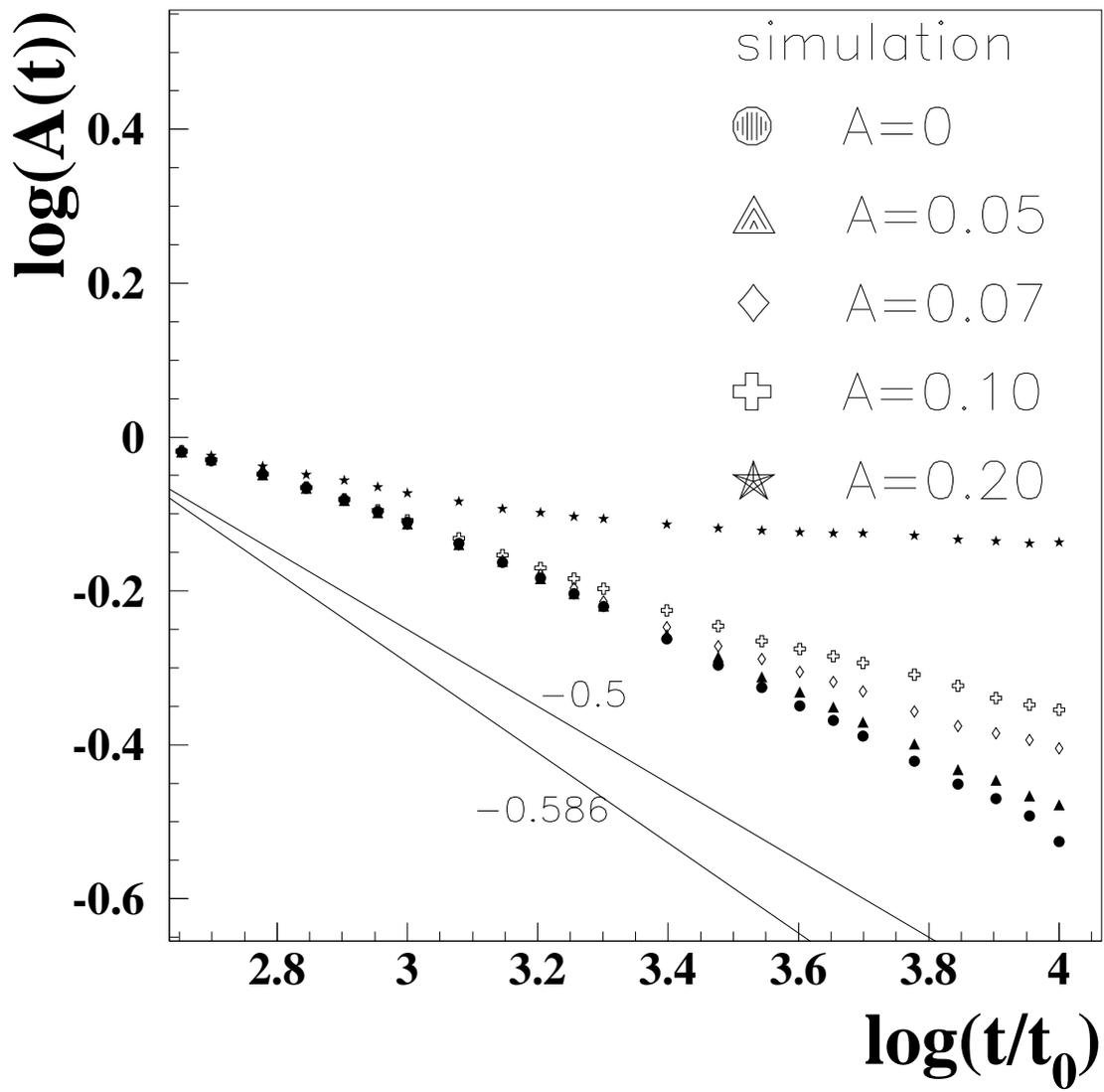
\noindent\underline{Figure 6.} Autocorrelation function for the $2d$ $XY$
model. Simulation data from Pargellis et al \cite{PGY:94}.
The solid lines with slopes $-0.5$ and $-0.586$ are the asymptotic decays
predicted by the systematic approach and by Mazenko's approach
\cite{LM:92:vectors,BH:92}.

\medskip

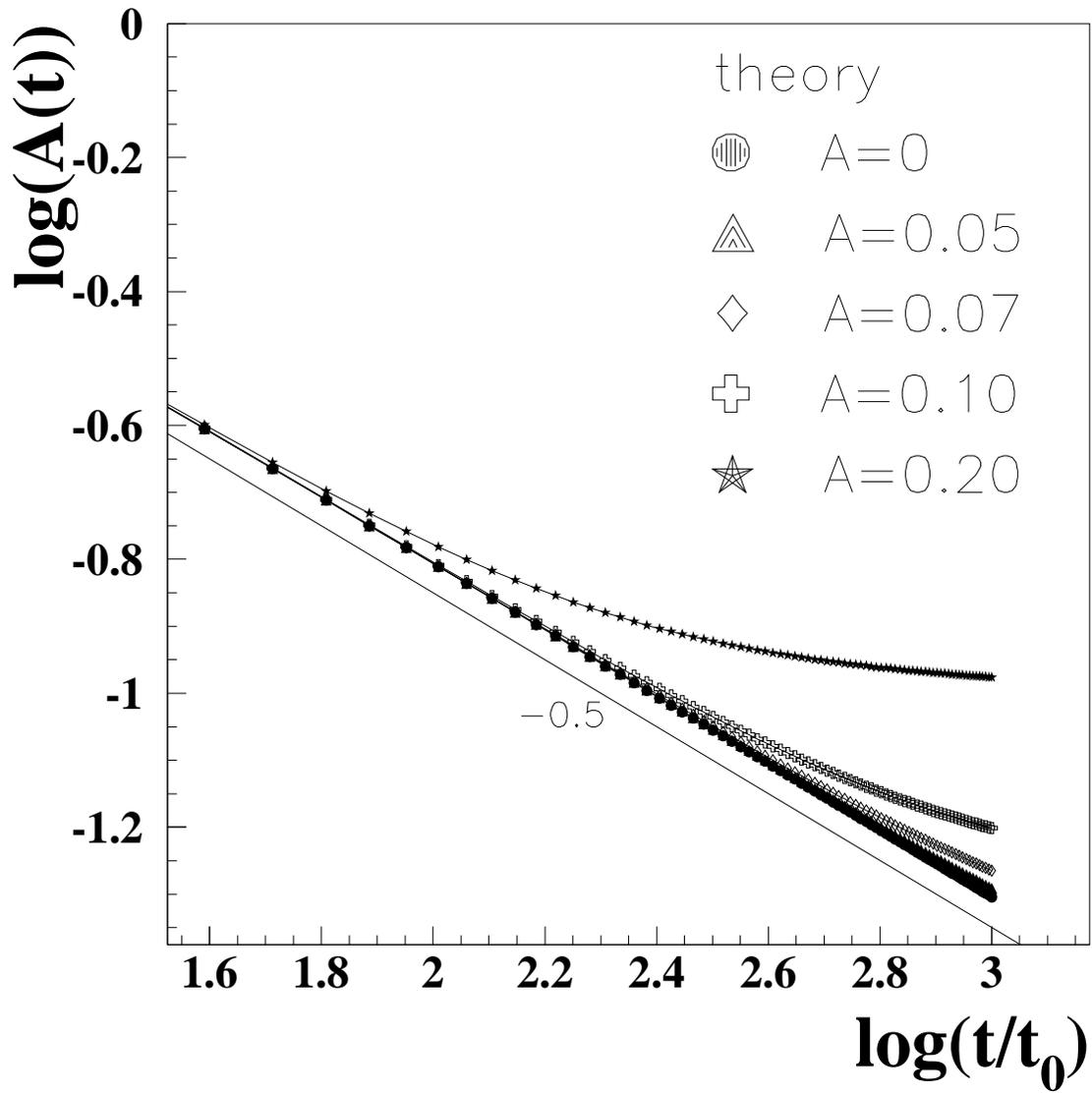
\noindent\underline{Figure 7.} Autocorrelation function for the $2d$
$XY$ model, as predicted by the systematic approach.

\medskip

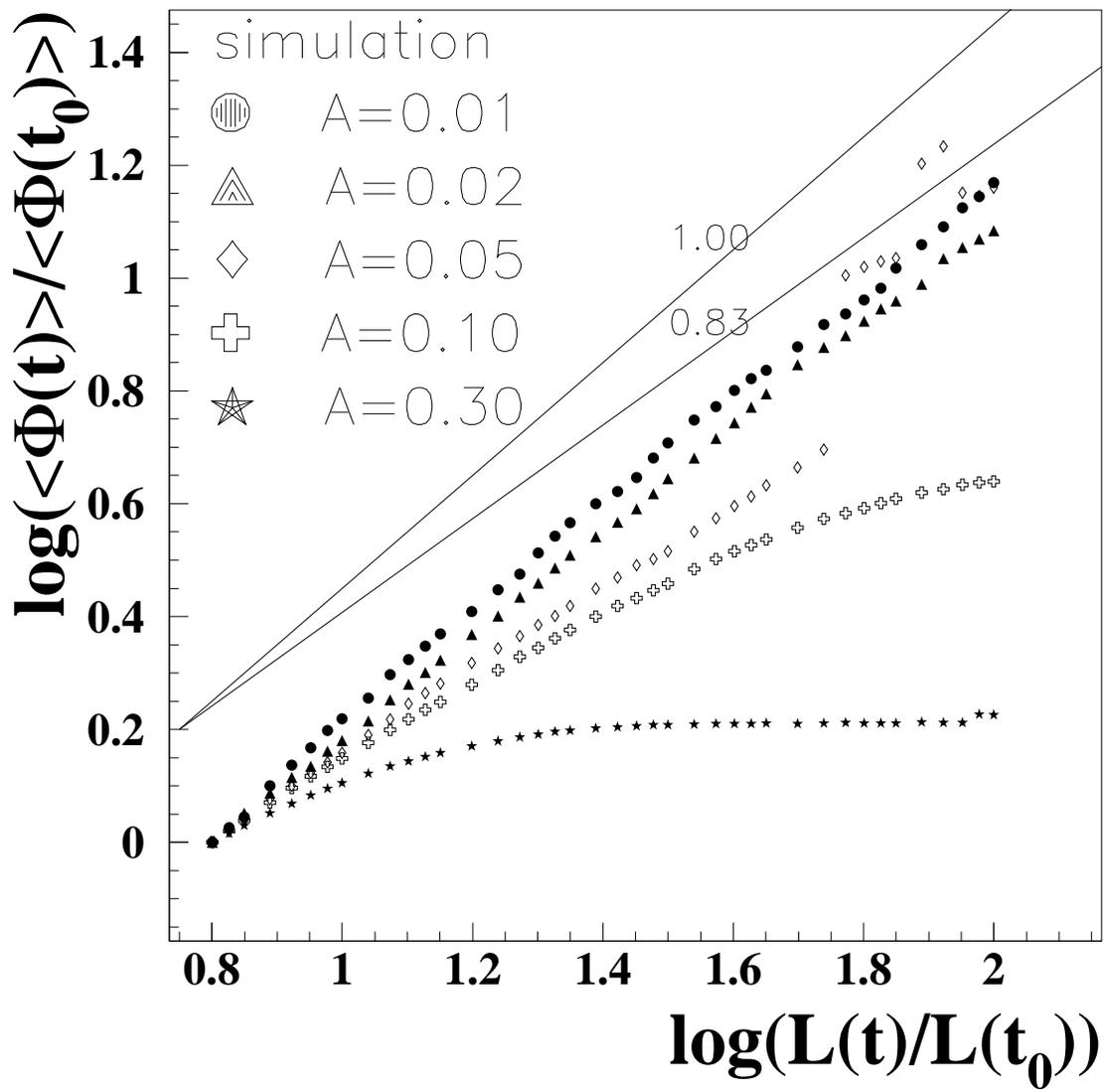
\noindent\underline{Figure 8.} Magnetization for the $2d$ $XY$ model.
Simulation data from Pargellis et al.\ \cite{PGY:94}
The solid lines with slopes $1$ and $0.83$ describe the initial growth as
predicted by the systematic approach and by Mazenko's approach
\cite{LM:92:vectors,BH:92}.

\medskip

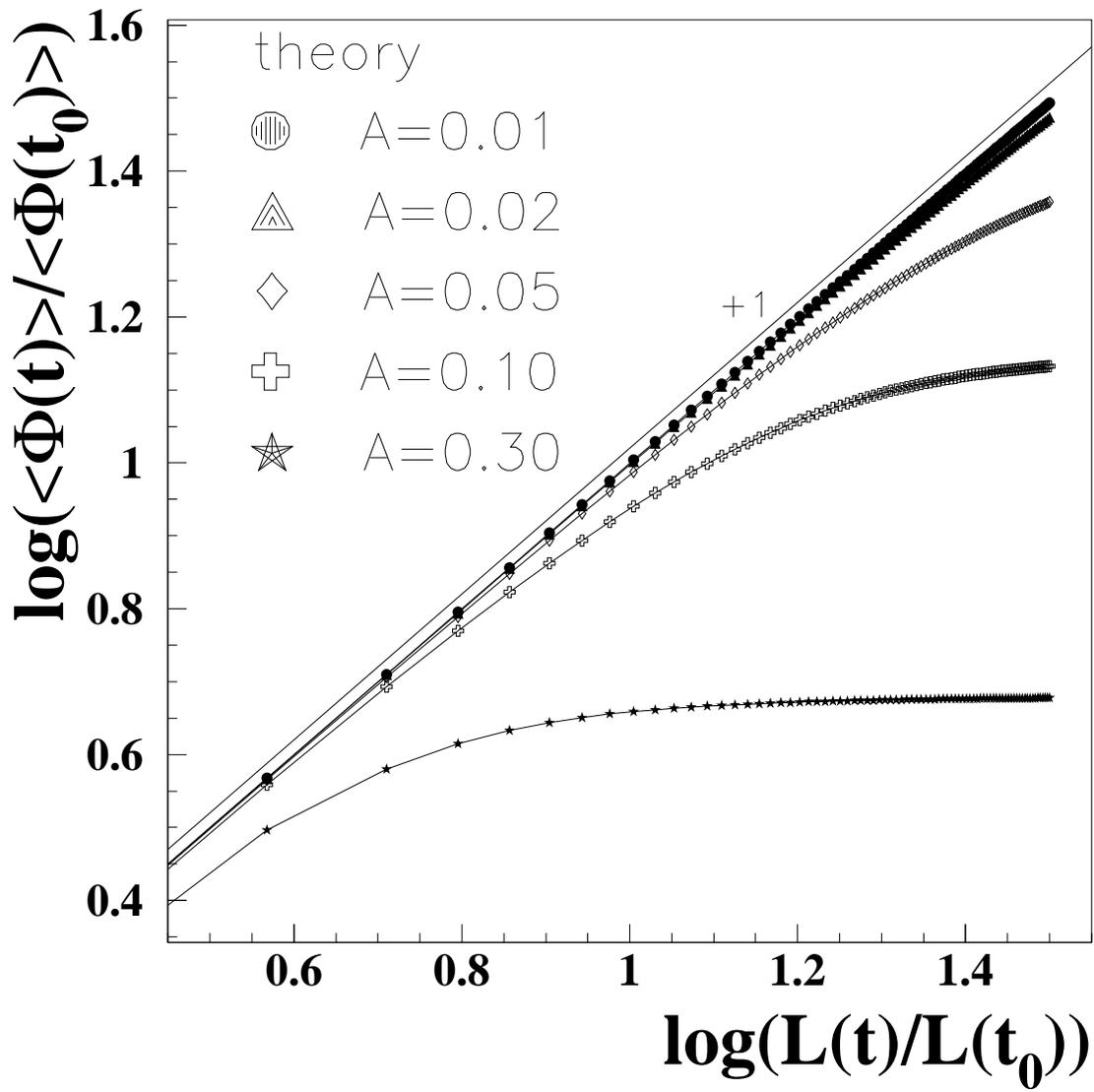
\noindent\underline{Figure 9.} Magnetization for the $2d$ $XY$ model,
as predicted by the systematic approach.


\begin{thebibliography}{99}
\bibitem{footnote} For example, for a spherical domain
with $\phi={\rm sgn}(h)$ and radius $R$ these forces (in the direction
of the centre) are $\Delta V/\Delta\phi \sim -h$ and
$(d-1)\Sigma/R$ respectively.
\end{thebibliography}
\end{document}